\documentclass{jfm} %[lineno]

\usepackage{graphicx}
\usepackage{epstopdf,epsfig}
\usepackage{newtxtext}
\usepackage{newtxmath}
\usepackage{natbib}
\usepackage{hyperref}
\hypersetup{
    colorlinks = true,
    urlcolor   = blue,
    citecolor  = black,
}

\usepackage[justification=justified]{caption}  %Line up the figure captions. 
\usepackage{textcomp}
\usepackage{amsmath}

\newcommand{\be}{\begin{equation}}
\newcommand{\ee}{\end{equation}}

\newcommand{\RomanNumeralCaps}[1]
\linenumbers

\title{Vortex Dynamics in Rotating Rayleigh-B\'enard Convection}

\author{Shan-Shan Ding\aff{1},
Guang-Yu Ding\aff{2,3},
Kai Leong Chong\aff{3,4},
Wen-Tao Wu\aff{1},
Ke-Qing Xia\aff{2,3} \corresp{\email{xiakq@sustech.edu.cn}} and
Jin-Qiang Zhong\aff{1,5} \corresp{\email{jinqiang@fudan.edu.cn}}}

\affiliation{\aff{1}School of Physics Science and Engineering, Tongji University, Shanghai 200092, China 
\aff{2}Center for Complex Flows and Soft Matter Research and Department of Mechanics and Aerospace Engineering, Southern University of Science and Technology, Shenzhen 518055, China
\aff{3}Department of Physics, The Chinese University of Hong Kong, Shatin, Hong Kong, China
\aff{4}Shanghai Institute of Applied Mathematics and Mechanics, School of Mechanics and Engineering Science, Shanghai University, Shanghai 200072, China
\aff{5}Department of Aeronautics and Astronautics, Fudan University, Shanghai 200433, China}

\date{\today}

\begin{document} 
\maketitle

\begin{abstract}

We investigate the spatial distribution and dynamics of the vortices in rotating Rayleigh-B\'enard convection in a reduced Rayleigh-number range $1.3{\le}Ra/Ra_{c}{\le}166$. Under slow rotations ($Ra{\gtrsim}10Ra_{c}$), the vortices are randomly distributed. The size-distribution of the Voronoi cells of the vortex centers is well described by the standard $\Gamma$ distribution. In this flow regime the vortices exhibit Brownian-type horizontal motion.
%with their mean-square-displacement (MSD) increasing in time as $t^2$ in the ballistic regime, and linearly in the diffusive regime. 
The probability density functions of the vortex displacements are, however, non-Gaussian at short time scales. 
At modest rotating rates ($4Ra_{c}{\le}Ra{\lesssim}10Ra_{c}$) the centrifugal force leads to radial vortex motions, i.e., warm cyclones (cold anticyclones) moving towards (outward from) the rotation axis. The mean-square-displacements of the vortices increase faster than linearly at large time. This super-diffusive behavior can be satisfactorily explained by a Langevin model incorporating the centrifugal force. In the rapidly rotating regime ($1.6Ra_{c}{\le}Ra{\le}4Ra_{c}$) the vortices are densely distributed, with the size-distribution of their Voronoi cells differing significantly from the standard $\Gamma$ distribution.    
The hydrodynamic interaction of neighboring vortices results in formation of vortex clusters. Inside clusters the correlation of the vortex velocity fluctuations is scale free, with the correlation length being approximately $30\%$ of the cluster length. 
We examine the influence of cluster forming on the dynamics of individual vortex. Within clusters, cyclones exhibit inverse-centrifugal motion as they submit to the motion of strong anticyclones, while the velocity for outward motion of the anticyclones is increased.
Our analysis show that the mobility of isolated vortices, scaled by their vorticity strength, is a simple power function of the Froude number.
%Our studies bring new perspectives into understanding the vortex dynamics in rotating convection that are widely present in nature.        

\end{abstract}

%{\bf MSC Codes }  {\it(Optional)} Please enter your MSC Codes here
%\input{1-Introduction}\label{sec:1}

\section{Introduction}

% Columnar vortex flow is ubiquitous in nature system, e.g., the coherent structure generating magnetic field in the Earth$'$s liquid core \citep{Jones11}, hurricanes \citep{Palmen48}, dust vortices on Mars and on earth \citep{Thomas85} \citep{Balme06}. However, the dynamics of columnar vortices are not well explored. Rotational Rayleigh-B\'enard convection (RRBC), the flow with applied vertical temperature difference and rotating around its central axis at a constant rate, is abundant of columnar vortices and offers an excellent platform for studying vortex dynamics.The non-dimensional parameters describing a cylinder system include the aspect ratio $AR$ describing the geometry of the cell, Rayleigh number describing the strength of thermal forcing, Prandtl number for the properties of working fluid, Ekman number for the rotating effect and Froude number for the strength of centrifugal force,

Buoyancy-driven convection is relevant to many natural flows in the atmosphere, oceans and planetary systems \citep{MS99,Va06,Jo11}. A rich variety of vortex structures arise during buoyant convection, especially in the presence of background rotations \citep{HV93}. Vortices are often referred to as coherent structures that consist of recirculating flows with roughly circular streamlines. The dynamics of vortices plays an important role in determining fluid motions and turbulent transport, ranging from small-scale turbulence to planetary-scale circulations \citep{FS01}. The dynamics convective vortices can be studied using a paradigmatic model, rotating Rayleigh-B\'enard convection (RBC), i.e., a fluid layer heated from below and rotated about a vertical axis. In rotating RBC, the fluid flows are governed by a sets of non-dimensional parameters including the Rayleigh number ($Ra$) describing the strength of thermal forcing, the Ekman number ($Ek$) representing the rotating effect, the Froude number ($Fr$) for the strength of centrifugal force, the Prandtl number ($Pr$) for the fluid properties and the aspect ratio $\Gamma_a$ describing the geometry of the fluid domain: 
 
\begin{equation*}
Ra=\frac{g\alpha \Delta T H^3}{\kappa \nu}, Ek=\frac{\nu}{2\Omega^2H}, Fr=\frac{\Omega^2D_{0}}{2g}, Pr=\frac{\nu}{\kappa}, \Gamma_a=D_{0}/H.
\end{equation*}
Here $g$ denotes the gravity acceleration, $\Delta T$ is the applied temperature difference, $D_{0}$ and $H$ are the horizontal and vertical length scales of the fluid domain, respectively.  $\alpha$, $\kappa$ and $\nu$ are the thermal expansion, thermal diffusivity, and kinetic viscosity of the working fluid. $\Omega$ is the applied rotating velocity along a vertical axis. 

A number of previous investigations report that for rapidly rotating RBC, the convective flows are organized by the Coriolis force into columnar vortices \citep{BG86, Sa97, VE02, PKVM08, KSNHA09, KCG10, GJWK10, JRGK12, SLDZ20}. The formation of a columnar vortex can be described by the theory of thermal wind balance, which states that vertical velocity gradients is caused by horizontal temperature gradients in the flow field \citep{KC08}
\begin{equation}
2\Omega \frac{\partial \vec{u}}{\partial z}=-\alpha g \hat{e_z}\times \nabla T. 
\label{eq:tb}
\end{equation}
When observed in the lower fluid level, upwelling warm fluid elements rotate in the same direction as the system when they experience horizontal advection, forming cyclones. Oppositely, cold downwelling fluid elements are organized into anticyclones. Under thermal wind balance, the horizontal gradient in the right hand side of Eq. (1.1) alters the vertical velocity and vorticity in magnitude along vertical axis and their signs around the middle plane. 
%Therefore, the vorticity strength of cyclones should be always larger than anticyclones independent of observation height. (It is not relevant, and unimportant) 
Both types of vortices are tall, thin, coherent convection columns with their horizontal scale given by $l{=}(2{\pi}^{4})^{{1/6}}Ek^{1/3}H$ for high-Pr fluids in the limit of rapid rotations \citep{Ch61, JK98, ABG18}. The number density $n$ of the columnar vortices increases when $Ek$ decreases. It is found in previous studies that $n$ can be described by a power-function of $Ek$, i.e., $n{\propto}Ek^{\gamma_n}$ \citep{BG86, Sa97, VE02, KCG10}. Different scaling exponents $\gamma_n$ varying from ${-}1.5$ to ${-}0.4$ are reported in these studies, which is presumably ascribed to the different fluid heights where measurements are made and the various criterion used to identify vortices. The mean spacing between the columnar vortices is given by $d_{v} \propto \sqrt{1/n}$ \citep{Sa97}, assuming that the vortices are uniformly distributed. However, when the spatial distribution of the vortices are not uniform, as observed by  \cite{Paper1} and \cite{Paper2} in rapidly rotating RBC, the vortex spacing $d_{v}$ is not simply a square-root function of $n$, then a more comprehensive description of the vortex spatial distribution is required. 

The flow structure of the columnar vortices in rotating RBC was investigated in numerical simulation \citep{JLMW96} and in laboratory experiments \citep{VE98}. These studies discovered that each columnar vortex is surrounded by a shielding layer. The signs of vorticity, temperature anomaly and vertical velocity in the shield layer are opposite to that in the vortex core region. \cite{PKVM08} proposed a theoretical model for the flow structures of the columnar vortices, in which they consider linearized governing equations of fluid motion and provide analytical solutions for the radial profiles of temperature fluctuations, vertical velocity and vorticity. 
An asymptotic theory of rapidly rotating RBC was developed by \cite{SJKW06,GJWK10}, where
they suggested that in the limit of extremely rapid rotations the columnar vortex structure is steady and axially and vertically symmetric, and predicted that the poloidal stream function of the vortices can be described by the zeroth-order Bessel function of the first kind \citep{GJWK10}. It follows that both the radial profiles of the azimuthal velocity and the vertical vorticity can be expressed by prescribed Bessel functions. These predictions appeared to match with numerical simulations \citep{GJWK10,NRJ14}. Recently, \cite{SLDZ20} performed spatially resolved measurements of the fine structures of the columnar vortices. Their results reveal that the asymptotic theory predicts accurately the velocity and vorticity profiles of the vortices in the flow regime of rotation-dominated convection, but deviate from the experimental results in weakly rotating convection. Three-dimensional experimental vorticity structure of vortices was measured by \cite{FTYNM20} through scanning velocity fields at different heights.

Numerical simulations in the full parameter space of rotating convection have revealed distinct flow structures with increasing buoyancy forcing, namely, cellular convection, convective Taylor columns, plumes, and geostrophic-turbulence \citep{JRGK12, SLJVCRKA14, NRJ14}. It is shown that in the flow regime of CTC the shield structure of the columnar vortices weakens with increasing $Ra$ and finally disappears. The violent vortex interactions then destroy the coherent columns and lead to the formation of plume-like structures \citep{SLJVCRKA14}. \cite{RKC17}performed measurements of the spatial vorticity autocorrelations to reveal the periodicity of the flow structures. Their experimental results indicated a sharp change in the slope of the correlation function, indicating a transition from the cellular-columnar state to the plume state. \cite{SLDZ20} measured the vorticity gradient at the shielding layer to determine quantitatively the strength of the vortex shield structures. They showed that the mean vorticity gradient followed two distinct scaling relations with increasing $Ra$, and suggested that a flow-state transition from weakly rotating convection to rotation-dominated, geostrophic convection.

%% vortex motion
%Although the large amount of work above reveals the vortex static characteristics, vortex dynamics has been less studied. \citet{Boubnov86},\citet{Sakai97} and \citet{Vorobieff02} have observed vortex motion in laboratory. \citet{Ecke93} observed vortex pair merging. \citet{Chong20} found that vortices exhibit normal diffusion both in laboratory and direct numerical simulation (DNS) when the centrifugal force is too weak to influence the vortex dynamics. They revealed the diffusion motion of the vortex is pure Brownian motion since the autocorrelation function of vortices' velocity decays with time in an exponential way. In fast RRBC, the influence of centrifugal force on flow dynamics can not be avoid. When the $Fr$ number is smaller than $0.5AR$ \citep{Horn18}, the geostrophic regime remains with the centrifugal force only acting as an perturbation to the dominant flows, i.e., columnar vortices. In this regime of fast RRBC, \citet{Noto19} found the centrifugal effect on vortex dynamics that the vortices exhibit super diffusion for a long time. \citet{Ding21} found cyclonic vortices exhibit inverse centrifugal motion which is caused by collective motion. The collective motion of vortices show similar properties with biological groups. 

Despite the large amount of studies devoted to explore the flow structures and spatial distribution of the vortices in rotating RBC, fewer efforts have been made to inspect the dynamics these vortices. Early experiment observations suggest that these convective vortices exhibit diverse dynamical states of horizontal motion, ranging from quasi-stationary, vortex merging to intensive advection \citep{BG86, ZES93, Sa97, VE02, KA12}. \cite{Paper1} demonstrate through both experiment and numerical simulation that the vortices undergo horizontally diffusive motion and resembles that of inertial Brownian particles, i.e., they move ballistically in short time but then diffusively in a large time scale. They reported that the diffusion motion of the vortices is in the type of pure Brownian motion, since the vortex velocity autocorrelation function decays exponentially in time, and the transition from ballistic to diffusive motion is sharp. Under rapid rotations the centrifugal force plays a role to influence the vortex motion. \cite{NTYM19} observed radial acceleration of the vortices that yielded a super-diffusive vortex motion in large time scale. \cite{HHXX21} used a convection cell placed at a distance away from the rotation axis in which they can vary the Froude number with a fixed Ekman to study the centrifugal effects. They reported an onset of flow bifurcation above which the cold and hot vortices moved in opposite directions with cold (hot) vortices concentrated in the far (near) region. \cite{Paper2} performed statistical analysis of the centrifugal motions of the convective vortices. They reported in the centrifugation-dominated flow regime the counterintuitive effect of hot vortices moving outward from the rotation axis, driven by long-range vortex interactions. Recently, \cite{DZCZ22} reported that when periodic topographic structures are constructed on the heated boundary in rotating RBC, the stochastic translational motion of the columnar vortices can be strictly controlled to form stationary convection patterns with prescribed symmetries. 

In this paper, we investigate the spatial distribution and dynamics of the convective vortices in rotating RBC. Our study covers the flow states from weakly rotating convection to rotation-dominated convection.
%where the vortices are sparsely or densely distributed. 
The structure of the paper is organized as following. We introduce our experimental and numerical methods in~\S~2. Discussions of the vortex spatial distribution will be presented in~\S~3. We describe the random diffusion and centrifugal motion of single vortex in~\S~4, and provide an extended Langevin model to interpret the vortex dynamics. %The dependence of vortex dynamics on system parameters will be given in~\S~\ref{sec:5}. 
Inverse-centrifugal motion of cyclonic vortices are reported in~\S~5, where we give details of the different flow regimes for vortex motions.
We discuss the dynamics of clustered vortices in the inverse-centrifugal flow regime in~\S~6. A summary and discussion of the results is provided in~\S~7.

\section{Experimental and Numerical Methods}\label{sec:2}

\subsection{Experimental setup and parameters}

The experimental apparatus was designed for high-resolution flow field measurements in rotating RBC. 
\citep{SLDZ20, Paper1, Paper2}. 
Here we present only its essential features. The bottom plate of the convection cell was made of oxygen-free, high-conductivity copper (OFHC). 
% It had a thickness of 35.0 mm and a diameter of 285.7 mm. 
Its bottom side was covered uniformly by parallel straight grooves. A main heater made of resistance wires was epoxied into the grooves. Seven thermistors were installed into the bottom plate, at a vertical distance of 3.5 mm from the top surface of the plate. The main heater was operated in a digital feedback loop in conjunction with these thermistors to hold the bottom-plate temperature as a constant with a stability of a few milli-Kelvin.  
For the purpose of flow visualization we used a top plate made by a 5 mm thick sapphire disc. A thermal bath that contained circulating coolant was constructed over the sapphire plate and regulated its temperature. During the experiment the temperature fluctuation of top plate was within 0.005 $K$.

The temperature difference between the bottom plate and the ambient air may induce thermal perturbations to the experiment. To eliminate mostly this temperature difference, a bottom adiabatic shield was installed under the bottom plate. This thermal shield was covered by a bottom-shield heater, with a thermistor located at the center of the shield.  A second auxiliary heater was wound around the periphery of the shield and the local temperature there was measured by a second thermistor. Both heaters worked in conjunction with their relevant thermistors to maintain the bottom-shield temperature the same as the bottom-plate temperature. 

In between the top- and bottom-plate was a cylindrical sidewall made of plexiglas in thickness of 3 mm. Thermal protection towards the sidewall was provided by a separate thermal side shield. It was a thin cylindrical ring made of aluminum, with a spiral aluminum tube wound on its outer surface. A circulating flow of coolant passed through the aluminum tube kept the side-shield temperature at the mean fluid temperature. Most of the spatial volume in between the thermal shields and the convection cell was filled with low-density open-pore foam to prevent convective air flows.
The aforementioned two coolant circuits were brought into the rotary table through a rotary union. It was a four-passage feed through equipped with a slip ring for electrical leads. The rotary table rotated clockwise driven by an electric servo-motor. All components of the convection cell were installed on the rotary table. 

%provided by FIBRO (model NC 1.04)

In the present study we used two cylindrical cells with an inner diameter $D_{0}{=}240$ mm, and fluid height $H{=}$63.0 (120.0) mm, yielding the aspect ratio $\Gamma_{a}{=}D_{0}/H{=}$3.8 (2.0). We report here mainly measurements in the cell with $\Gamma_{a}{=}3.8$ unless otherwise noted. Deionized water was the working fluid with a constant Prandtl number ${Pr}{=}\nu/{\kappa}{=}4.38$. The experiment was conducted in the range $2.0{\times}10^7{\le}Ra{\le}2.7{\times}10^8$ of the Rayleigh number ${Ra}{=}{\alpha}g{\Delta}TH^3/{\kappa}{\nu}$ ($\alpha$ is the isobaric thermal expansion coefficient, $g$ the acceleration of gravity, $\Delta T$ the applied temperature difference, $\kappa$ the thermal diffusivity and $\nu$ the kinematic viscosity). All measurements were made at constant $\Delta T$ with $\Omega$ varying from 0 to 4.7 rad/s. The Ekman number ${Ek}{=}\nu/2{\Omega}H^2$ spanned $1.7{\times}10^{-5}{\le}Ek{\le}2.7{\times}10^{-4}$. 
%Thus the reduced Rayleigh number $\mathrm{\tilde{Ra}}{=}\mathrm{RaEk}^{4/3}$, which characterizes the relative strength of rotation, covered the range $10{\le}\mathrm{\tilde{Ra}}{\le}412$. 
The Froude number ${Fr}{=}{\Omega}^2D_{0}/2g$ was within $0{<}Fr{\le}0.31$.

We conducted measurements of the horizontal velocity field ($u_x, u_y$) using a particle image velocimetry system installed on the rotary table. A thin light-sheet power by a solid-state laser illuminated the seed particles in a horizontal plane at a fluid height $z{=}H/4$. Images of the particle were captured through the top sapphire window by a high-resolution camera (2456$\times$2058 pixels). Two-dimensional velocity fields were extracted by cross-correlating two consecutive particle images. Each velocity vector was calculated from an interrogation windows (32$\times$32 pixels), with $50\%$ overlap of neighboring sub-windows to ensure sufficient accuracy \citep{SLDZ20,WEA13}. Thus we obtained 154$\times$129 velocity vectors on each frame, reaching a resolution of 2.0 mm for the velocity field. We identify the vortices through a two-dimensional Q-criterion. Considering the quantity $Q{=}(Tr \mathbf{A})^2{-}4{det} \mathbf{A}$ with the velocity gradient tensor $\mathbf{A}{=}[\partial(u_x,u_y)/\partial(x,y)]$, we define the vortex center as the minimum of $Q$ within a vortex region satisfying $Q{<}{-}Q_\mathrm{std}$. Here $Q_\mathrm{std}$ is the standard deviation of $Q$ over the measured area $r{\le}D_{0}/4$ \citep{Paper1, Paper2}. We adopt the method of vortex tracking introduced in \citep{Paper2} to obtain vortex trajectories. 

\subsection{Numerical method}
In the direct numerical simulation (DNS) we solved the three-dimensional Navier-Stokes equations within the Boussinesq approximation: 
\begin{equation} 
%\begin{split}
%\frac{D\vec{u}}{Dt}=&-{\nabla}P+(\frac{\mathrm{Pr}}{\mathrm{Ra}})^{1/2}{\nabla}^2\vec{u}+{\theta}\hat{z} \\
%&+(\frac{\mathrm{Pr}}{\mathrm{RaEk}^2})^{1/2}\vec{u}{\times}\hat{z}-\frac{2r\mathrm{Fr}}{\tilde{D}}{\theta}\hat{r}, 
%\end{split}
\frac{D\vec{u}}{Dt}=-{\nabla}P+(\frac{Pr}{Ra})^{1/2}{\nabla}^2\vec{u}+{\theta}\hat{z}+(\frac{Pr}{RaEk^2})^{1/2}\vec{u}{\times}\hat{z}-\frac{2rFr}{D}{\theta}\hat{r}, 
\end{equation}
\begin{equation} 
\frac{D{\theta}}{Dt}=\frac{1}{(RaPr)^{1/2}}{\nabla}^2{\theta},
\end{equation}
\begin{equation} 
{\nabla}{\cdot}\vec{u}=0.
\end{equation}
Here $\vec{u}$ is the fluid velocity, $\theta$ and $P$ are the reduced temperature and pressure. 
%The last two terms in the momentum equation (Eq.\ 1) represent the Coriolis force and the centrifugal force. 

Equations (2.1)-(2.3) were nondimensionalized using $H$, ${\Delta}T$ and the free-fall velocity $U_{f}{=}\sqrt{\alpha g\Delta T H}$. The top (bottom) plate was isothermal with temperature $\theta_{t}{=}{-}0.5$ ($\theta_{b}{=}0.5$), and the sidewall was thermally insulated. As for the momentum boundary condition, all boundaries were non-slip. These equations were solved using the multiple-resolution version of {\it{CUPS}} \citep{KX13,CDX18}, which was a fully parallelized direct numerical simulation code based on finite volume method with 4th order precision. To improve computational efficiency without any sacrifice in precision, we used a multiple-resolution strategy, i.e., the momentum equation was solved in a coarser grid than the temperature one, allowing for a sufficient resolution to resolve the Batchelor and Kolmogorov length scales. The simulations were performed in a cylindrical domain with $\Gamma_{a}{=}4$, ${Ra}{=}2.0{\times}10^{7}$, ${Pr}{=}{\nu}/{\kappa}{=}4.38$ and $1.3{\le}{Ra/Ra_c}{\le}55$.

\section{Vortex Spatial Distribution}\label{sec:3}

In the rotating RBC, the spatial distribution and organization of the convective vortices depend sensitively on the strength of rotation and buoyancy. To demonstrate the distribution distribution of the columnar vortices, we present in figure 1 Voronoi diagrams of the vortex centers for two reduced Rayleigh numbers, $Ra/Ra_{c}{=}8.90$ and 2.67, respectively. The background color of these diagrams represents distribution of the $Q$ field, where vortices are shown in reddish color and the bluish areas in-between indicate regions of high flow strain. We see that under weak rotations the vortices appears randomly located with a broad size distribution for $Ra/Ra_{c}{=}8.90$. The Voronoi cells becomes smaller in average but with an even size-distribution for rapid rotations ($Ra/Ra_{c}{=}2.67$).

\begin{figure}
\centerline{\includegraphics[width=0.95\textwidth]{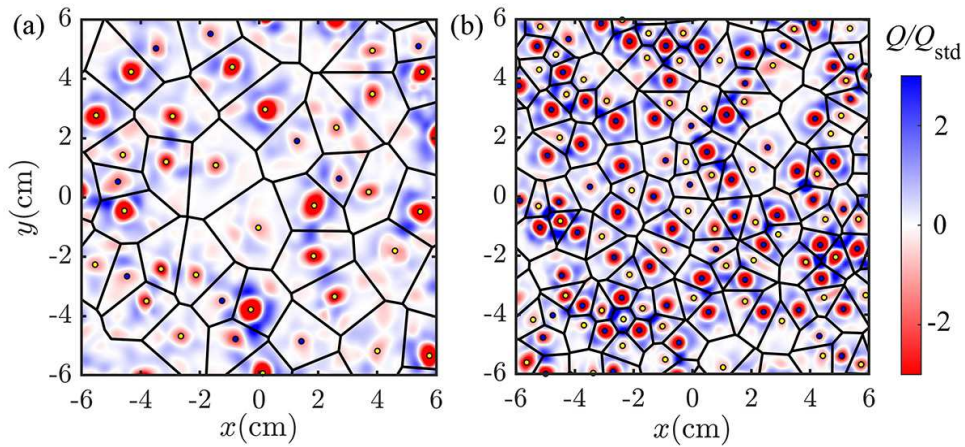}}% Images in 100% size
% \caption{Voronoi diagram generated from centers of vortices for $Ek=5.37\times 10^{-5}$ (a) and $2.24\times 10^{-5}$ (b) with $Ra=3\times 10^7$ and $AR=3.8$. The background color represents $Q/Q_{\rm std}$ and black(yellow) dots represent the centres of anticyclones(cyclones). The solid lines denote edges of voronoi cells.}
\caption{Spatial distribution of the vortices for weakly rotating convection (a) and rotation-dominated convection (b). The solid-line network constitutes the Voronoi diagram of the vortex centers. The background coloration represents distribution of the quantity $Q/Q_{std}$, 
%that reveals the strength of the vorticity field, 
with black (yellow) dots denoting the centers of anticyclones (cyclones). Results for $Ra{=}3.0{\times}10^7$ and for $Ra/Ra_{c}{=}8.90$ (a) and  $Ra/Ra_{c}{=}2.67$ (b).}
\label{fig:voronoi}
\end{figure}

We calculate the probability density functions (PDFs) of Voronoi cell area normalized by the mean value $A/{\langle}A{\rangle}$. It has been shown that for randomly distributed entities in $d$-dimension space ($d{\le}3$), the PDF of the scaled areas (volumes) of the Voronoi cells is a standard $\Gamma$-distribution with the dimension $d$ as the only fitting parameter: 

\begin {equation}    
{
P(x)=\frac{[(3d+1)/2]^{(3d+1)/2}}{\Gamma[(3d+1)/2]}{x}^{\frac{3d-1}{2}}\mathrm{e}^{-[(3d+1)/2]x}
}
\label{eq:gamma}
\end{equation}
\noindent

Here the denominator is a $\Gamma$ function. For two-dimensional distributions we have 
$P_2(x){=}7^{7/2}2^{-7/2}x^{5/2}e^{-7x/2}/{\Gamma}(7/2)$. 
Figure 2a shows that for a sufficiently large $Ra/Ra_{c}$, $P(A/{\langle}A{\rangle})$ follows closely the standard $\Gamma$-distribution $P_2(x)$. With the increasing rotation rate (decreasing $Ra/Ra_{c}$), the PDFs of the Voronoi cell area show different behavior from that of the standard $\Gamma$-distribution. We see that the PDFs have a smaller possibility for very large and small cells than for random distributed one, whereas a larger probability for $A/{\langle}A{\rangle}{\approx}1$. These results suggest that the vortices are indeed randomly distributed for $Ra/Ra_{c}{=}83.09$ and have a trend of forming regular vortex lattice when $Ra/Ra_{c}$ decreases.   

Figures 2b and 2c present the mean value ${\langle}A{\rangle}$ and the variance $\sigma(A/{\langle}A{\rangle})$ of the Voronoi cell area as functions of $Ra/Ra_{c}$. Both our experimental and numerical data show that ${\langle}A{\rangle}$ decreases monotonically with decreasing$Ra/Ra_{c}$, signifying an increasing vortex number density. The variance $\sigma(A/{\langle}A{\rangle})$, however, exhibits the interesting trend that it first decreases till reaching a minimum value at $Ra/Ra_{c}{\approx}5$. When $Ra/Ra_{c}{\le}5$, $\sigma(A/{\langle}A{\rangle})$ starts to increase. The emergence of the minimum of $\sigma(A/{\langle}A{\rangle})$ at $Ra/Ra_{c}{\approx}5$ can be seen also in figure 2a. The decreasing of $\sigma(A/{\langle}A{\rangle})$ for $Ra/Ra_{c}{\ge}5$ implies  the homogenized size-distribution of the Voronoi cells with decreasing $Ra/Ra_{c}$. For $Ra/Ra_{c}{\le}5$, however, the increase of $\sigma(A/{\langle}A{\rangle})$ is ascribed to the rapid decreasing of ${\langle}A{\rangle}$ with decreasing $Ra/Ra_{c}$ as shown in figure 2b.   

\begin{figure}
\centerline{\includegraphics[width=0.95\textwidth]{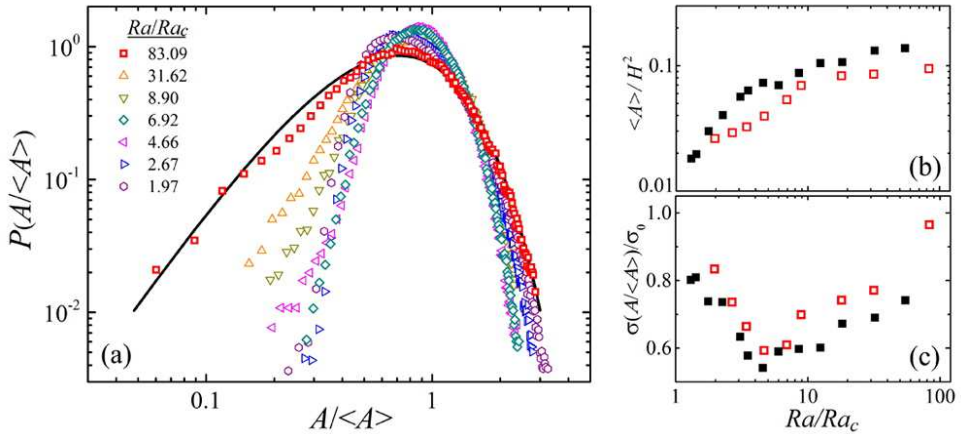}}
  \caption{ (a) PDFs of the Voronoi cell area $P(A/{\langle}A{\rangle})$ for various $Ra/Ra_{c}$. The solid curve represents the two-dimensional $\Gamma$-distribution $P_2(A/{\langle}A{\rangle})$. (b) The normalized mean area $\left<A\right>/H^{2}$ of the Voronoi cells as a function of $Ra/Ra_{c}$. (c) The rescaled standard deviation $\sigma(A/{\langle}A{\rangle})/\sigma_0$ as a function of $Ra/Ra_{c}$, with $\sigma_0{=}\sqrt{2/(3d{+}1)}{=}\sqrt{2/7}$. Open symbols: experimental data for $Ra{=}3.0{\times}10^7$. Closed symbols: numerical data for $Ra{=}2.0{\times}10^7$.}
\label{fig:vodistr}
\end{figure}

\section{Single Vortex Dynamics}\label{sec:4}
\noindent

We discuss in this section the rich dynamics of translational motion of sparsely distributed vortices observed in a horizontal plane, including Brownian-type random diffusion in the flow regime of weakly rotating convection, and centrifugal motion with stochastic fluctuations in the centrifugation-influenced convection regime. Some of our results have been reported briefly before \citep{Paper1,Paper2}. Here we present a full description and provide more comprehensive discussions of these two types of stochastic vortex motion.

\subsection {Randomly diffusive motion of vortices in weakly rotating convection}

\begin{figure}
	\centerline{\includegraphics[width=1.0\textwidth]{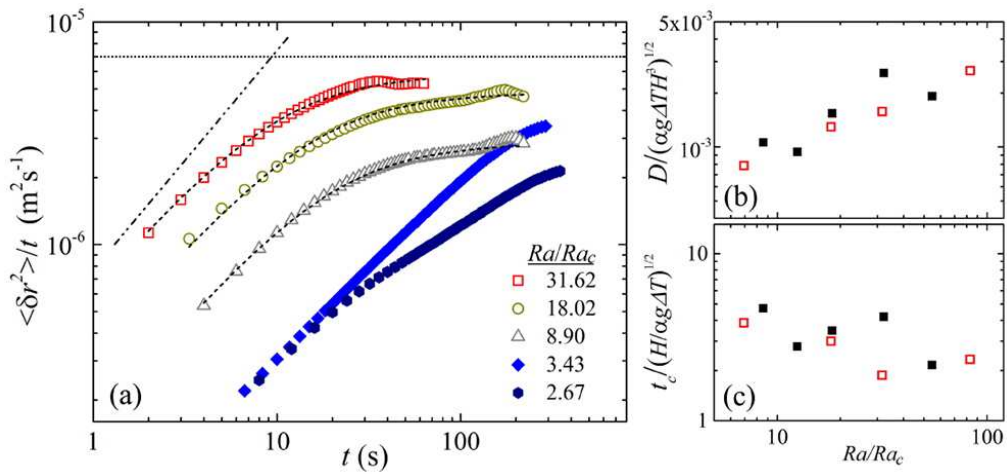}}
	\caption{(a) Results of $\langle \delta r^2\rangle/t$ as a function of $t$ for various $Ra/Ra_c$. The dashed lines present the theoretical fits to the data for weakly rotating convection based on (4.2). The dash-dot line indicates a scaling $\langle \delta r^2\rangle{\sim}t^{2}$ in the ballistic regime. Results for $Ra{=}3.0{\times}10^7$. (b) and (c) show the normalized diffusivity $D$ and transitional time $t_{c}$ as a function of $Ra/Ra_c$. 
		% \textcolor[rgb]{1,0,0}{Here we have performed data fitting using (4.2) in a parameter range where the centrifugal effect is present but not strong enough to have notable influence on the statistics of $\langle \delta r^2\rangle$.} 
	Open symbols: experimental data for $Ra{=}3.0{\times}10^7$. Closed symbols: numerical data for $Ra{=}2.0{\times}10^7$.}
	\label{fig:msd}
\end{figure}

Through tracking the horizontal motion of the vortices, we obtain their trajectories $\vec{r}(t)$ of the vortices, and calculate the mean-square-displacement (MSD) for the vortices for a given time interval $t$, i.e., $\langle\delta r^2\rangle(t){=}\langle \left(\vec{r}(t){-}\vec{r}(0)\right)^2\rangle$. Figure 3 shows $\langle\delta r^2\rangle/t$ as a function of $t$ for several $Ra/Ra_{c}$ in the full parameter range. We first discuss here results in weakly rotating convection, where the centrifugal effect is minor. It is clearly see that in this flow state the vortex motion undergoes a transition from the ballistic regime where $\langle\delta r^2\rangle$ increases as $t^2$ in short time, to a diffusive regime where $\langle\delta r^2\rangle$ increases linearly and $\langle\delta r^2\rangle/t$ approaches a constant in long time. Thus the vortex motion resembles the random motion of Brownian particles, which can be described by the Langevin Equation:
\begin{equation}
\ddot r+ {\dot r}/{t_c}={\xi}(t),
\label{eq:bmw}
\end{equation}
where the two momentum terms on the left-hand-side are the inertia and the viscous force, respectively. ${\xi}(t)$ represents the stochastic force acting on the vortices that is modeled as ${\langle}\xi(t){\rangle}{=}0$ and ${\langle}\xi(t)\xi(t{+}{\Delta}t){\rangle}{=}q{\delta}({\Delta}t)$. Here $q$ represents the strength of the stochastic force and   
% which is modeled as a stochastic force with white-noise spectrum:  
%\begin{equation} 
%\label{eq:noisedef}
%{\langle}\xi(t){\rangle}=0,   {\langle}\xi(t)\xi(t+{\Delta}t){\rangle} = {\delta}({\Delta}t)D/\tau^2,
%\end{equation}
$\delta(t)$ is a Dirac function. Solving Eq. 4.1 one obtains the solution for the MSD of the vortices: 
%\noindent
\begin{equation}
\langle \delta r^2\rangle = 2Dt[1-t_c/t(1-\mathrm{e}^{-t/t_c})].
\label{eq:solbmw}
\end{equation}
Here $t_c$ is the transitional time from the ballistic to diffusive regime. It represents the characteristic time scale of relaxation for the diffusive vortex motion. $D$ is the diffusivity that relates to the noise strength through $D{=}qt_{c}^{2}$ according to the fluctuation-dissipation theorem.
The dashed lines in figure 3a are fitting curves $\langle \delta r^2\rangle/t$ to the experimental data according to (4.2), with two fitting parameters ($t_c, D$) that depends on $Ra/Ra_c$. Figures 3b and 3c present results of $t_c$ and $D$ as functions of $Ra/Ra_c$. Both our experimental and numerical data show that with decreasing $Ra/Ra_c$,  $t_c$ increases while $D$ decreases monotonically. The increasing relaxation time $t_c$ when $Ra/Ra_c$ decreases is not yet well understood, but we infer that in the flow regime of weakly rotating convection the vertical scale of the vortices grows when the rotating rate $\Omega$ increases since the vortex structure has not fully developed to penetrate the whole fluid layer \citep{SLDZ20}. The larger vertical scale of the vortices results in a greater inertia of motion and thus an increasing relaxation time $t_c$ for the translational vortex motion. The apparent decreasing in $D$ signifies the reducing intensity of the stochastic force acting on the vortices when $Ra/Ra_c$ decreases.  

% signifying the reduction of the random force on the vortices.
% need more discussion on inertia

\begin{figure}
	\centerline{\includegraphics[width=1.0\textwidth]{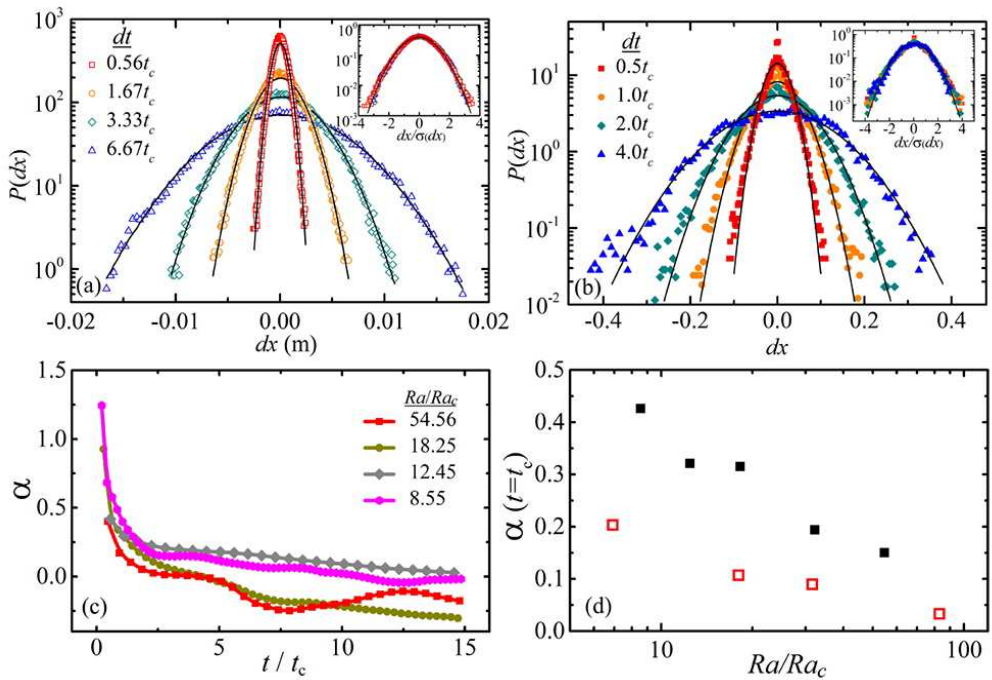}}
	\caption{(a) and (b) PDFs of the vortex displacement $P(dx)$ at various time intervals $dt$. Experimental (a) and numerical (b) data for $Ra/Ra_{c}{=}18.0$. Insets in (a) and (b) show the rescaled PDFs $P(dx/{\sigma}(dx))$ where the vortex displacement $dx$ is normalized by its standard deviation ${\sigma}(dx)$. (c) The non-Gaussian parameter $\alpha$ as a function of $t/t_c$. (d) $\alpha$ as a function of $Ra/Ra_{c}$ for $t=t_c$. Open symbols: experimental data for $Ra{=}3.0{\times}10^7$. Closed symbols: numerical data for $Ra{=}2.0{\times}10^7$.}
	\label{fig:pdfdx}
\end{figure}

To gain deep insight into the statistical properties of the vortex motion in the state of weakly rotating convection, we study the PDFs of the vortex displacement $P(dx,dt)$ in one dimension $dx$ for different time intervals $dt$. Figures 4a and 4b show experimental and numerical data for $Ra/Ra_{c}{=}18.0$, respectively. We see that for a large time interval $dt{\ge}t_{c}$, $P(dx)$ is Gaussian to good precision, as one would expect for normal Brownian motion. For very small time interval $dt{\le}t_{c}$, however, $P(dx)$ appears to deviate from a Gaussian function. In order to investigate the deviation of $P(dx)$ from Gaussian distribution with decreasing $dt$, we calculate the excess kurtosis $\alpha$, which characterizes the departure from Gaussianity, $\alpha{=}
{\langle}(dx{-}{\langle}dx{\rangle})^4{\rangle}/[3{\langle}(dx{-}{\langle}dx{\rangle})^2{\rangle}^2]
{-}1$. Thus $\alpha{=}0$ indicates a perfect Gaussian distribution and a large $\alpha$ signifies a departure from Gaussian. Figure.\ 4c shows results of $\alpha$ as a function of $dt/t_c$ for two sets of $Ra/Ra_c$. One sees that $\alpha{\le}0.1$ for large time interval $dt{\ge}t_{c}$. However, when $dt$ decreases below $t_{c}$ $\alpha$ increases rapidly and becomes significant for very small $dt$. Results of the MSD and the PDFs of the vortex displacement (figures 3 and 4) suggests that the at very short time scales the vortex motion undergoes non-Gaussian yet Brownian diffusion. A full theoretical interpretation of the diffusive dynamics of the vortices remains yet to be studied. 

%Here we infer that at a very short time scale the stochastic force acting to drive the vortex motion, which originates mainly from the background turbulent flows, may be correlated in space and time rather than simply white noise. The observed PDFs of the vortex displacement in this time scale is thus non-Gaussian. As the time interval $dt$ increases and so does the vortex displacement $dx$, the vortices are subject to the large-amplitude forcing owing to the complex vortex-vortex interaction. As the vortices in this weakly rotating regime follows a $\Gamma$-type, random spatial distribution (figure 2), we hypothesize that the vortex interaction in a large time scale is random which dominates the turbulent fluctuation and results in a Gaussian-type, normal vortex diffusion. 

Here we infer that in a small time scale the random motion of the vortices, driven by the turbulent fluctuations of the background flows, is largely disturbed by the passing vortices. The vortex interactions, such as merging (annihilation) of same(opposite)-sign neighboring vortices, results in intermittent but deterministic horizontal movements of the vortices in short time. The observed PDFs of the vortex displacement in this time scale is thus non-Gaussian. 
As the time interval $dt$ increases and so does the vortex displacement $dx$, the intermittent 
perturbation from single adjacent vortex gives way to the complex interactions from multiple neighboring vortices that appear to have stochastic spatiotemporal properties. In addition, owing to the large-amplitude background turbulent fluctuations that dominate the vortex motion in large time scales, the PDF of the vortex displacements return to Gaussianity.
In figure 4d we investigate the non-Gaussian parameter $\alpha$ as a function of $Ra/Ra_c$ for a given time interval $dt{=}t_{c}$. Both our experimental and numerical data suggest that in this weakly rotating regime $\alpha$ increases when $Ra/Ra_c$ decreases. A similar trend of increasing $\alpha$ with decreasing $Ra/Ra_c$ is found when other time intervals $dt$ are chosen. 

%an interesting trend that with decreasing $Ra/Ra_c$, $\alpha$ first increases and then start to decrease when $Ra/Ra_c{\le}6.9$. The fact that the maximum of $\alpha$ appears at $Ra/Ra_c{=}6.9{\pm}1.0$ irrespective of $dt$ remains to be further understood in future studies.                 

\subsection {Stochastic motion of vortices driven by centrifugal force}

Figure 3a shows that in the state of rapidly rotating convection the MSD $\langle\delta r^2\rangle(t)$ of the vortices increases faster than a linear function of $t$ for large time, which implies super-diffusive behavior of the vortices in rapidly rotating convection. This phenomenon is attributed to the centrifugal effect (See. e.g. \cite{NTYM19,HHXX21,HXX22}): in a reference frame of rotation warm (cold) vortices are driven radially inwards (outwards) by the centrifugal force. Such radial motions add to the displacement of the vortices and result in their superdiffusion in large time. In the parameter range considered in our study, such a centrifugal effect acts as an external forcing that is capable to alter the horizontal motion of the vortices, but not as strong to modify their coherent flow structures \citep{HA18,HA19}.  

\begin{figure}
  \centerline{\includegraphics[width=0.8\textwidth]{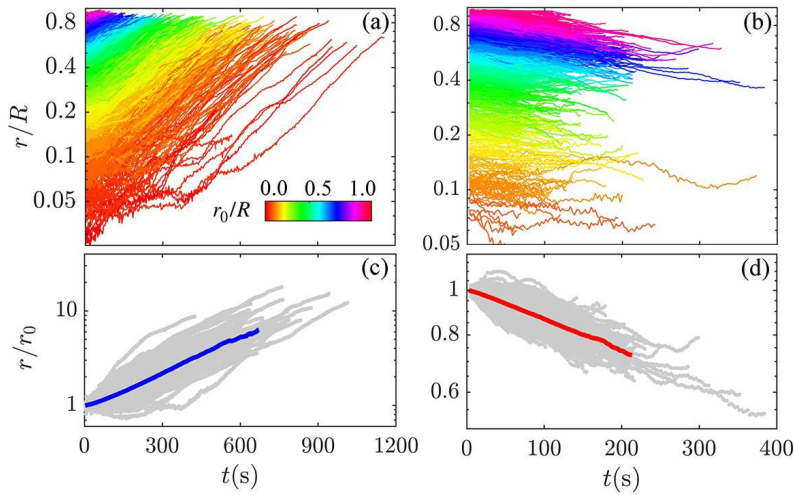}}
  \caption{Space-time plots of representative radial trajectories of the vortices. (a) and (b) show data of $r(t)/R$ for anticyclones and cyclones, respectively. The color represents the initial radial position $r_0/R$ of the vortices. (c) and (d) show $r(t)/r_{0}$ for anticyclones and cyclones. The blue line in (c) and the red line in (d) represent the ensemble-averaged trajectories $\langle r/r_0\rangle_{\xi}(t)$. In (b) and (d), data for cyclones that exhibit inward motions are shown to illustrate normal centrifugal effects. Data are for $Ra{=}3.0{\times}10^7$, $Ra/Ra_{\mathrm c}{=}1.97$ and $Fr{=}0.27$.}
\label{fig:traj}
\end{figure}

Figures 5a and 5b are space-time plots of representative radial trajectories of anticyclones and cyclone, respectively.  The radial positions of the vortices are shown as functions of time for various initial positions $r_0$. It appears that the radial displacements $r(t)$ of anticyclones (cyclones) increase (decrease) approximately exponentially in time. Figures 5c and 5d present the scaled radial trajectories of the vortices $r(t)/r_0$, corresponding to the same data sets in figures 5a and 5b, respectively. We then compute the ensemble-averaged scaled radial trajectories $\langle r/r_0\rangle_{\xi}$, which is the average of $r(t)/r_0$ of all trajectories for a given time $t$. Results of $\langle r/r_0\rangle_{\xi}$(t) are shown by the thick blue (red) lines for anticyclones (cyclones), which indicate clearly the exponential growth (decay) of the vortex radial displacements. 

In order to understand the super-diffusive behavior of the vortices in a background of stochastic movements, we propose a model to describe the radial vortex motion under the centrifugal force. Our model has been introduced before \citep{Paper2}. For the purpose of integrity for the present work, we provide here brief discussions of its derivation and predictions. We consider Langevin-type equations of motion for the vortices, i.e., $m{\ddot{r}}{+}{\eta}{\dot{r}}{\pm}m{\alpha}{\delta}T{\Omega}^2r{=}{\xi}^{\ast}(t)$. The vortices are subject to the centrifugal force expressed in terms of ${\pm}m{\alpha}{\delta T}{\Omega}^2 r$ for warm cyclones (plus sign) and cold anticyclones (minus sign). Here $m$ is inertia mass of the vortices and ${\delta}T$ is the temperature difference between the vortex and the background fluid. We define the relaxation time $t_c{=}m/\eta$ and the centrifugal coefficient $\zeta{=}{\alpha}{\delta T}{\Omega}^2$ to obtain: 
\begin{equation}
\ddot r+{\dot r} / t_c \pm \zeta r={\xi}(t)
\label{eq:scenf}
\end{equation}
Here ${\xi}{=}{\xi}^{\ast}(t)/m$ is the rescaled strength of background turbulent fluctuation, which modeled by a ${\delta}$-correlated white noise ${\langle}\xi(t)\xi(t{+}{\Delta}t){\rangle}{=}{\delta}({\Delta}t)D/{t_{c}}^2$. $D$ is the diffusivity. Defining valuables $\lambda_{1,2}{=}\pm \sqrt{{-}\zeta{+}{1}/({4{t_c}^2}}){+}{1}/({2{t_c}})$ for cyclones and  $\lambda_{1,2}{=}\pm \sqrt{\zeta{+}{1}/({4{t_c}^2}}){+}{1}/({2{t_c}})$ for anticyclones, one derives from (4.3) the solution for the first moment of radial displacement:
\begin{equation}
\langle r/r_0\rangle_\xi=\frac{\lambda_1}{\lambda_1-\lambda_2}\mathrm{e}^{-\lambda_2t}-\frac{\lambda_2}{\lambda_1-\lambda_2}\mathrm{e}^{-\lambda_1t}
 \label{eq:fm}
\end{equation}
where $\langle ... \rangle_\xi$ presents average over all trajectories of the vortices. In the large-time limit ($t \gg t_{\mathrm c}$) the first moment of radial displacement asymptotes to a single exponential function, 
${\langle}r/r_0{\rangle}_{\xi}{\approx}{\lambda_1}\mathrm{e}^{\pm \lambda^*t}/({\lambda_1{-}\lambda_2})$. Here $\lambda^{*}{=}|1/(2t_c){-}\sqrt{1/({4t_c}^2)\pm \zeta}|$ is the fastest growth (or slowest decay) rate for anticyclones (cyclones), which represents the mobility of vortices in the context of centrifugal acceleration. One thus derive the asymptotic solution of the ensemble-averaged velocity in large time,
\begin{equation}
\langle u_r/r_0 \rangle _{\xi} \approx \frac{\lambda_1\lambda^*t}{\lambda_1-\lambda_2} \mathrm{e}^{\pm \lambda^*t} \approx \lambda^*\langle r/r_0 \rangle _{\xi} 
\label{eq:ur}
\end{equation}
where $u_r$ is the radial velocity of vortices.

We can further obtain the second moment of the radial displacement as follows:
%further derived to study the stochastic properties in vortex radial motion, that is,
\begin{equation}
{\langle}\left[r(t){-}\langle r(t)\rangle_\xi\right]^2\rangle_\xi =
\frac{D}{{t_c}^2(\lambda_2-\lambda_1)^2}\left[\frac{1-{\mathrm{e}}^{-2{\lambda_1}t}}{2{\lambda_1}}+\frac{1-{\mathrm{e}}^{-2{\lambda_2}t}}{2{\lambda_2}}-\frac{2-2{\mathrm{e}}^{-{(\lambda_1+\lambda_2)}t}}{{\lambda_1}+\lambda_2}\right]
 \label{eq:sm}
\end{equation}

The first moment $\langle r/r_0 \rangle _{\xi}$ represents the mean radial displacement of the vortices, while the second moment ${\langle}\left[r(t){-}\langle r(t)\rangle_\xi\right]^2\rangle_\xi$ measures the standard deviation of the radial vortex displacements from the mean (see figures 5c and 5d for demonstrations). As reported elsewhere \citep{Paper2}, our theoretical predictions (4.4) and (4.6) of these statistical quantities are in close agreements with the experimental data. Our model provides satisfactorily explanations of the super-diffusive behavior of the convective vortices observed in rotating RBC systems (e.g. \cite{NTYM19, Paper2}).

\section{Anomalous Regime of Vortex Motion}\label{sec:5}

\begin{figure}
  \centerline{\includegraphics[width=1.0\textwidth]{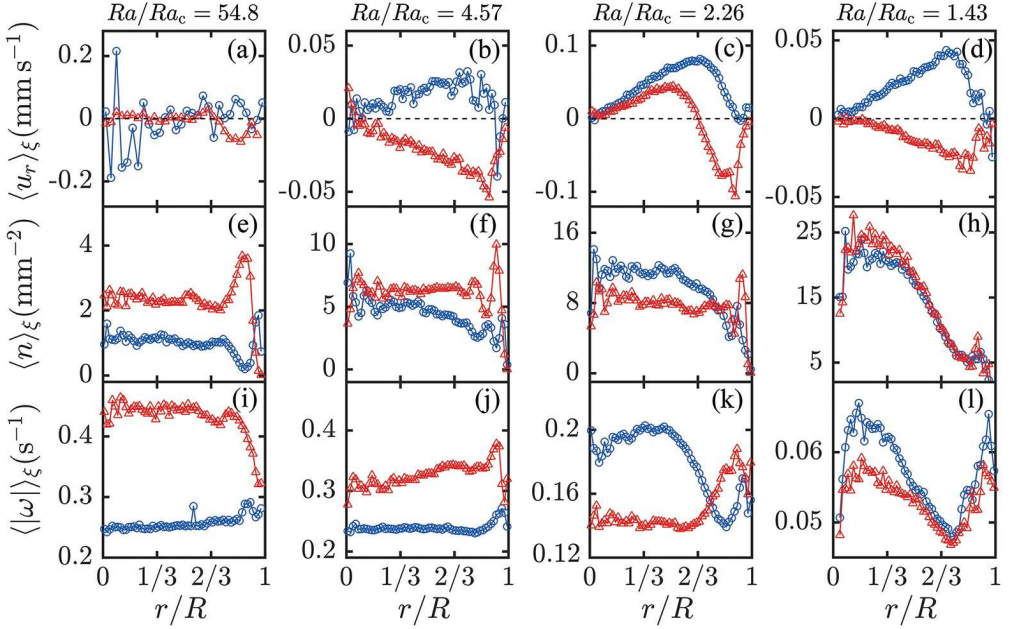}}
  \caption{The ensemble-averaged radial velocity $\langle u_r \rangle _\xi$ (a-d), vortex density $\langle n\rangle_{\xi}$ (e-h) and vorticity amplitude $\langle|\omega|\rangle_{\xi}$ (i-l) as functions of $r/R$. $\left< \right>_{\xi}$ denotes the ensemble average.
Results are for $Ra{=}2.0{\times}10^7$ with $Ra/Ra_{c}{=}54.8$ and $Fr{=}0.030$ (a,e,i), $Ra/Ra_{c}{=}4.57$ and $Fr{=}0.044$ (b,f,j), $Ra/Ra_{c}{=}2.26$ and $Fr{=}0.12$ (c,g,k), $Ra/Ra_{c}{=}1.43$ and $Fr{=}0.24$ (d,h,l), respectively. Red triangles (blue circles) denote cyclonic (anticyclonic) data.}
\label{fig:unw}
\end{figure}

The centrifugal force gives rise to the radial vortex motion, resulting in radial-dependence of the vorticity field. Figure 6 shows as functions of the radial position, the mean values of the radial velocity $\langle u_r \rangle _\xi$, number density $\langle n\rangle_{\xi}$ and vorticity magnitude $\langle|\omega|\rangle_{\xi}$ of both types of vortices. Four distinct flow regimes are clearly identified. We first find in a weakly rotating regime ($Ra/Ra_c{=}54.8$ and $Fr{=}0.030$) that $\langle u_r \rangle _\xi$ for both types of vortices fluctuates around zero, indicating the absence of radial motion (figure 6a). In this flow regime, we note that the mean values of number density $\langle n\rangle_{\xi}$ and vorticity magnitude $\langle|\omega|\rangle_{\xi}$ for cyclones are significantly greater than those for anticyclones (figures 6e and 6i). This is the case because anticyclones are down-welling vortices generated from the top boundary. When observed at the lower half of the fluid layer, they travel a longer distance to the measured fluid height ($z{=}H/4$) than the upwelling vortices (cyclones), and their momentum and vorticity have been largely dissipated by the background turbulence. ($\langle n\rangle_{\xi}$ and $\langle|\omega|\rangle_{\xi}$ for the two types of vortices would be equal if measured at $z{=}H/2$ (see discussions in reference \citep{Paper2}). In the sidewall-region ($r/R{>}0.85$) large fluctuations of all the measured variables appear owing to the intensive perturbation of the flow field by the boundary zonal flows \citep{ZvHWZAEWBS20,WGMCCK20}. 

When the rotating rate increases ($Ra/Ra_c{=}4.57$, $Fr{=}0.044$), we see that $\langle u_r \rangle _\xi$ of anticyclones (cyclones) is larger (smaller) than zero indicating that anticyclones (cyclones) move away from (towards) the rotation axis under the centrifugal force. Figure 6b shows that $\langle u_r \rangle _\xi$ increases (or decreases) linearly with $r$ for anticyclones (cyclones), which is well predicted by our model (4.5). Cyclonic vortices still possess a greater number density and vorticity magnitude than those of anticyclones at the measurement height (figures 6f and 6j). With further increase of $\Omega$ the centrifugal effect becomes dominant. For $Ra/Ra_c{=}2.26$, $Fr{=}0.12$, we note that unexpectedly $\langle u_r \rangle _\xi$ of both vortices are positive and increase linearly with $r$ in the inner region (figure 6c), signifying that cyclones exhibit outward motion that is opposite to the centrifugal effect. For both types of vortices $\langle u_r \rangle _\xi$ reaches a maximum at a radial position that depends on $Ra/Ra_c$, and then decreases with larger $r$ in the outer region. We find in the central region both the number density and vorticity magnitude of anticyclones exceed those of the cyclones (figures 6g and 6k). In the limit of rapid rotation with $Ra/Ra_c{=}1.43$, $Fr{=}0.24$, the cyclones are found to take up the inward radial motion with $\langle u_r \rangle _\xi$ decreases linearly with $r$ in the inner region (figure 6d). The profiles of $\langle n\rangle_{\xi}(r)$ and $\langle|\omega|\rangle_{\xi}(r)$ for both types of vortices show similar radial-dependence, suggesting that the symmetry of the vorticity field restores near the onset of convection. 

\begin{figure}
  \centerline{\includegraphics[width=1\textwidth]{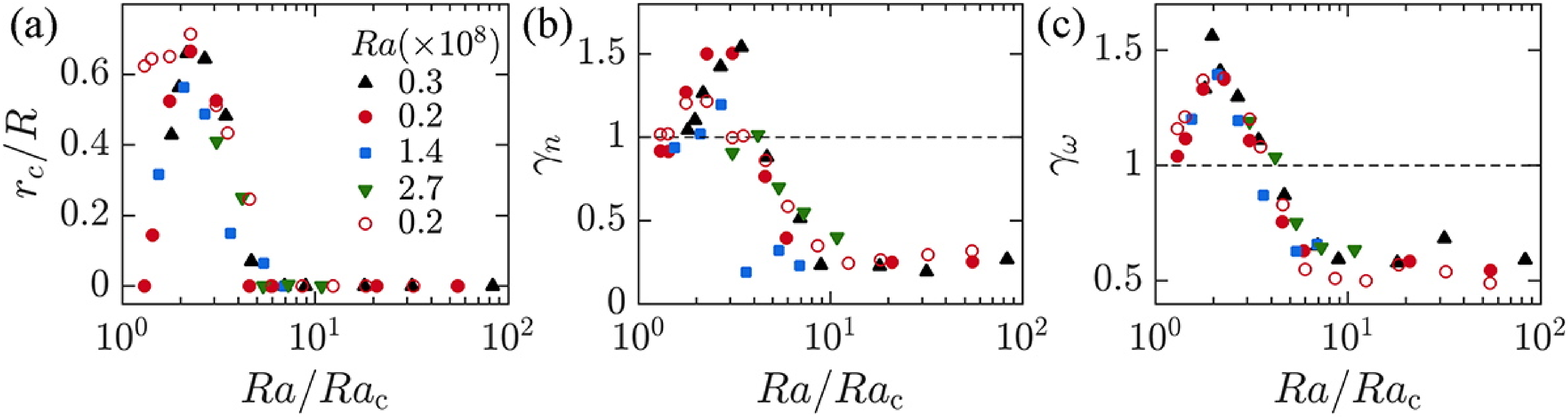}}
  \caption{ (a) The outer region boundary $r_c/R$ for the inverse centrifugal motion of cyclones. (b) The ratio $\gamma_n$ of the number density of anticyclones over cyclones. (c) The ratio  $\gamma_{\omega}$ of vorticity amplitude of anticyclones over cyclones. Results are shown as functions of $Ra/Ra_c$. Data for $Ra{=}2.0{\times}10^{7}$ and $\Gamma_{a}{=}3.8$ (circles), $Ra{=}3.0{\times}10^{7}$ and $\Gamma_{a}{=}3.8$ (up-triangles), $Ra{=}1.4{\times}10^{8}$ and $\Gamma_{a}{=}2.0$ (squares), $Ra{=}2.7{\times}10^{8}$ and $\Gamma_{a}{=}2.0$ (down-triangles). The open circles are numerical data.}
\label{fig:regime}
\end{figure}

We further investigate the parameter regime in which the inverse centrifugal motion of cyclones occurs. We define the region boundary $r_{c}$ for anomalous vortex motion as follows: the mean radial velocity of cyclones $\langle u_r \rangle _\xi{\ge}0$ for $r{\le}r_{c}$, and $\langle u_r \rangle _\xi{<}0$ for $r{>}r_{c}$. As shown in figure 6c, $r_{c}$ is thus the second zero crossing of the radial profile $\langle u_r \rangle _\xi(r)$. Figure 7a shows $r_{c}/R$ as a function of $Ra/Ra_c$. We find $r_{c}{=}0$ for $Ra/Ra_c{\ge}4$.  $r_{c}$ exceeds zero when $Ra/Ra_c$ decreases below 4, reaching a peak at $Ra/Ra_c{\approx}2$. 
With further decreasing in $Ra/Ra_c$ we see that $r_{c}$ decreases, and the experimental data indicate the trend that $r_{c}$ eventually approaches to zero near the onset of convection $Ra/Ra_c{\approx}1$. Near onset our numerical data show a higher value of $r_{c}$ than the experimental one, presumably owing to the insufficient numerical data evaluating the velocity profile $\langle u_r \rangle _\xi(r)$. Otherwise results of $r_{c}/R$ for various $Ra$ collapse approximately onto one single curve. We calculate in the central region ($r{\le}0.5R$) the ratios of the number density $\gamma_n$ and the vorticity magnitude $\gamma_{\omega}$ of the anticyclones over the cyclones. Figures 7b and 7c present results of $\gamma_n$ and $\gamma_{\omega}$ as functions of $Ra/Ra_c$, respectively. We see that data of $\gamma_n$ and $\gamma_{\omega}$ for various $Ra$ also exhibit similar dependence of $Ra/Ra_c$. For $Ra/Ra_c{>}4$, $\gamma_n$ and $\gamma_{\omega}$ are constants irrespective of $Ra/Ra_c$. $\gamma_n$ and $\gamma_{\omega}$ increase when $Ra/Ra_c{\le}4$ and have a maximum at $Ra/Ra_c{\approx}2$. When $Ra/Ra_c{<}2$ the two ratios $\gamma_n$ and $\gamma_{\omega}$ decrease with further decreasing $Ra/Ra_c$ and approaches unity at the onset of convection.  

The observed asymmetry of the vorticity field in the anomalous regime is attributed to the centrifugal effect. As the centrifugal force drives continuously hot (cold) fluid parcels towards (away from) the center of the convection cell, the background fluid temperature in the central region increases and exceeds global mean fluid temperature \citep{HO99, LE11,HA19}. Thus in the central region, the temperature difference of the cold anticyclones from the background fluid becomes greater than that of the warm cyclones. Since such a temperature anomaly is proportional to the buoyancy forcing on the vortices, it is positively correlated to the vorticity magnitude of the vortices \citep{PKVM08,GJWK10}. It is also believed that the fluid warming in the central region enhances the stability of the anticyclonic flows, leading to a larger population of anticyclones. As a results, we find in figures 7b and 7c that both the number density and the vorticity magnitude of the anticyclones exceeds the cyclonic ones.   

The aforementioned vortex dynamics and symmetric properties of the vorticity field reveal four distinct flow regimes depending on the rotation rates: (I) A randomly diffusive regime in the slow rotating limit with $Ra$ being one order in magnitude larger than $Ra_{c}$. In this flow regime the vortices move in a random manner, yielding $\langle u_r \rangle _\xi{\approx}0$, and $r_{c}/R$ is close to zero. Since in the measured fluid height the cyclones have a greater population as well as a larger vorticity magnitude than the anticyclones, $\gamma_n$ and $\gamma_{\omega}$ are both less than unity but independent of $Ra/Ra_{c}$. (II) A centrifugation-influenced regime ($4{\le}Ra/Ra_{c}{\lesssim}10$) where the magnitude of $\langle u_r \rangle _\xi$ increases linearly with $r$ (figure 6b). We observe that warm cyclones (cold anticyclones) move radially inward (outward), which is in agreement with the centrifugal effect. (III) An Inverse-centrifugal regime ($1.6{\le}Ra/Ra_{c}{\le}4$) in which the cyclones exhibits anomalous outward motion in the inner region with $r{\le}r_{c}$ (figure 6c). 
%We note that the radial gradients of $\langle u_r \rangle _\xi$ for both types of vortices are positive and reach a maximum at $Ra/Ra_{c}{\approx}2$. 
In this flow regime anticyclones become the dominant flow structures in the sense that both $\gamma_n$ and $\gamma_{\omega}$ exceed far above unity (figures 7b and 7c). In the outer region ($r{\ge}r_{c}$) the centrifugal effect of fluid warming becomes insignificant, we observe there inward cyclonic motion, as $\langle u_r \rangle _\xi$ for cyclones decreases with increasing $r$ and becomes negative. 
%(Note that in figures 6 and 7 we use the cyclonic data in the outer region with $r{\ge}r_{c}$ to determine the dynamical properties of their centrifugal motion.)
(IV) The asymptotic regime in the rapid rotation limit ($Ra/Ra_{c}{<}1.6$) where $r_{c}$ approaches zero and the opposite radial motions of cyclones and anticyclones recover (figure 6h).

\section{Vortex Cluster Dynamics}\label{sec:6}

\subsection{Formation of Vortex Clusters}

We have seen that in the inverse-centrifugal regime the symmetry of vorticity field is broken with the anticyclonic vortices dominating the cyclones both in strength and population. We show below that in this regime the vortices self-organize into clusters in which the anticyclones dominate the long-range correlated vortex motion, leading to the inverse centrifugal motion of the cyclones. 

\begin{figure}
  \centerline{\includegraphics[width=0.8\textwidth]{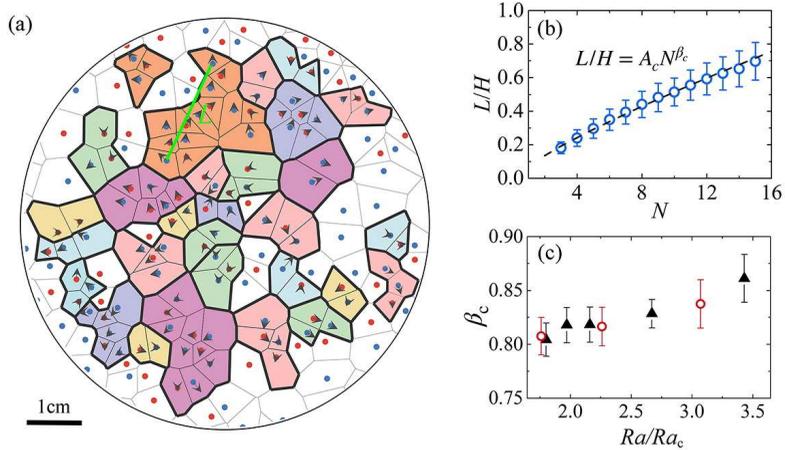}}
  \caption{(a) Spatial distribution of vortex clusters. The polygons represent the Voronoi tessellations of the vortex centers. Voronoi cells within the same vortex cluster are printed in the same color and enclosed by thick lines. The blue (red) dots denote the centroid of anticyclonic (cyclonic) vortices with arrows indicating their velocity direction. 
Experimental data in a region of $r{\le}0.7R$ are shown.  (b) The normalized cluster length $L/H$ as a function of cluster size $N$. The dashed line represents the fitted power function $L/H{=}A_{\mathrm c}N^{\beta_{\mathrm c}}$, where $A_{\mathrm c}{=}0.079$ and $\beta_{\mathrm c}{=}0.81$. Data in (a) and (b) are for $Ra{=}3.0{\times}10^7$ and $Ra/Ra_{c}{=}1.97$. (c) The exponent $\beta_{\mathrm c}$ as a function of $Ra/Ra_{c}$ for various $Ra$ numbers. Open circles:  $Ra{=}2.0{\times}10^7$. Solid triangles: $Ra{=}3.0{\times}10^7$.}
\label{fig:cluster}
\end{figure}

Figure 8a presents an example of instantaneous spatial distribution and motion of the vortices in the inverse-centrifugal regime. We find that the adjacent vortices often move in similar directions and aggregate locally, forming vortex clusters. Here we adopt the following two criteria to identify vortex clusters (See e.g. \cite{CDBSZ12}):

1) The distance of two neighboring vortices is smaller than 1.5 times the mean vortex diameter;

2) The angle between the velocity of two adjacent vortices is less than $\theta^{\ast}$.

Our analysis over the range $30^{\circ}{\le}\theta^{\ast}{\le}75^{\circ}$ confirms that the results of correlated vortex motion are not sensitive to the choice of $\theta^{\ast}$. In the following we discuss statistical properties using $\theta^{\ast}{=}60^{\circ}$.

In figure 8 vortex clusters are identified using the aforementioned criteria. In this Voronoi diagrams of vortex centers, each cluster is enclosed by thick lines, with Voronoi cells within the same cluster printed in the same color. We define the cluster length $L$ as the largest distance between two vortex centers within the cluster, and compute the number of vortices $N$ with each cluster. Figure 8b shows $L$ as a function of $N$ for $Ra{=}3{\times}10^{7}$ and $Ra/Ra_{c}{=}1.97$. We find that $L(N)$ can be well fitted by a power function, $L{=}A_cN^{\beta_c}$. Assuming that the distance between two adjacent vortices $d_{vv}$ is independent of $N$, the power exponent $\beta_c$ then reveals the dimensional properties of clusters. When $\beta_c{=}0.5$ vortices in a cluster are distributed isotropically in a two-dimensional plane. When $\beta_c{=}1$, vortices line up to form one-dimensional clusters. Figure 8c shows that within the full parameter range of $Ra/Ra_{c}$ studied, $\beta_c$ increases slightly from 0.80 to 0.87 with increasing $Ra/Ra_{c}$, suggesting a fractal property of the vortex clusters: vortices in a cluster have a weak preference to be arranged along one direction, while the horizontal span of the cluster is not isotropic but dependent on the orientation. 

%% cluster formation
\begin{figure}
 \centerline{\includegraphics[width=0.9\textwidth]{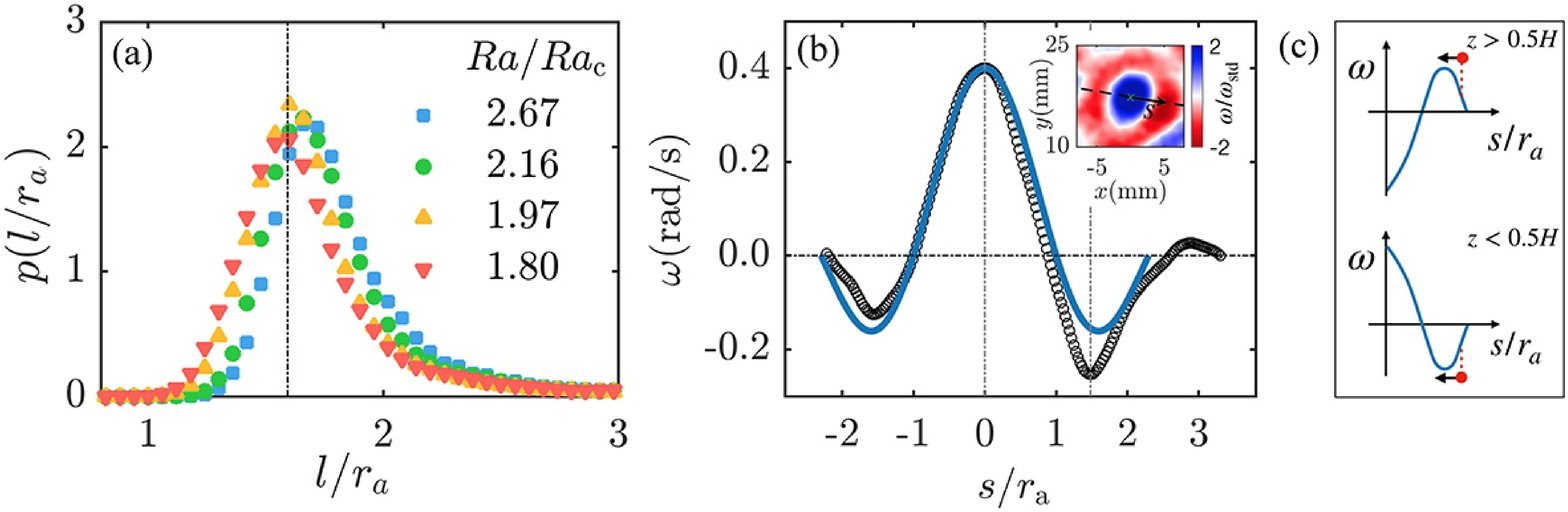}}
  \caption{(a) PDFs of the distance $l$ between an anticyclone and its neighboring cyclones within a cluster, normalized by the radius of the anticyclone $r_{\mathrm {a}}$. The vertical dashed line shows a maximum of $p(l/{r_{\mathrm{a}}})$ at $l{=}l_m{=}cr_{\mathrm {a}}$. (b) Vorticity profile $\omega(s/{r_{\mathrm{a}}})$ of an opposite-sign vortex pair along the centerline $s$. The anticyclonic radius $r_{\mathrm{a}}$ is defined by the radial position where $\omega(s/{r_{\mathrm{a}}})$ crosses zero. Open circles: experimental data. Solid line: the scaled zero-order Bessel functions: $J_0(k_s/{r_{\mathrm{a}}})\omega(0)/J_0(0)$, where $k{=}2.405$ is the first zero of $J_0$. The two vertical dashed lines indicate the centers of the anticyclone and the cyclone at $s_1{=}0$ and $s_2{=}cr_{\mathrm{a}}$. The inset shows the vorticity distribution of this vortex pair with the dashed line being the centerline. (c) Schematic plot of the interaction between a pair of opposite-signed convective columnar vortices at upper layer (top) and lower layer (bottom). The blue curve represents the vorticity profile of a downwelling vortex, which creates a background vorticity field that influences the motion of an adjacent upwelling vortex (denoted by the red circle).
Data are for $Ra{=}3.0{\times}10^7$ and $Fr{=}0.27$.}%(c) Histogram of the numbers of anticyclones (blue) and cyclones (red) that are contiguous to one cyclone. Insets: schematic of three clustering modes.  }
\label{fig:vpair}
\end{figure}

Within each cluster the vortices exhibits long-range correlated motions. We attribute these correlated vortex motions to the vortex-vortex interaction that occurs in a convection state where the vortices are densely distributed. Figure 9a shows the PDF $p(l/r_{a}$) of the distance $l$ between an downwelling vortex (anticyclone when observed in the lower half fluid layer) and its neighboring upwelling vortices for various $Ra/Ra_{c}$. Here $r_{a}$ is the radius of the downwelling vortex. One sees that $p(l)$ has an apparent maximum at $l_m{=}cr_a{=}1.593r_a$. Here the constant $c$ is the ratio of first minimum over the first zero of the zero-order Bessel function $J_0$. Since $p(l)$ represents the probability finding upwelling vortices at a distance $l$ from a downwelling vortex, it reflects the interactions between adjacent counter-rotating vortices. Figure 9b shows an example of the vorticity profile $\omega(s)$ of two opposite-sign, neighboring vortices, with the coordinate $s$ measuring the distance to the center of the downwelling vortex along the connecting line. We see that within the core region ($s{<}l_m$) the vorticity profile $\omega(s)$ is well described by $J_0(s)$. $\omega(s)$ reaches a first minimum at $s{=}l_m$ where $p(l)$ is maximum. Such a pair-wise vorticity profile is commonly observed within vortex clusters in our experiment. 

We provide the following interpretation for the most probable vortex separation $l_m$. We consider the motion of warm, upwelling vortices in the vicinity of a cold, downwelling vortex. Observed in the lower half fluid layer ($z{<}H/2$), the cold vortex gives rise to negative background vorticity gradient in the core region ($s{<}l_{m}$), but a positive vorticity gradient outside the vortex core ($s{>}l_{m}$) (see a schematic drawing in figure 9c). The theory of vortex motion on a vorticity gradient \citep{SD99} suggest that a upwelling vortex, which possesses negative vorticity in the lower half layer in our case, moves down the vorticity gradient due to the background shear flow. Thus, inside the core region ($s{<}l_{m}$) the upwelling vortex moves away from the downwelling vortex center, but moves towards it when $s{>}l_{m}$. Since a vortex with positive vorticity moves up a background vorticity gradient, the upwelling vortex undergoes the same translational motions in the upper half layer. As a result, we find $s{=}l_{m}$ the most probable radial position where an upwelling vortex locates as shown in figure 9a. For the same reason, one finds most probably a downwelling vortex at a radial distance $s{=}cr_c$ from an upwelling vortex center (with $r_c$ being the radius of the upwelling vortex). Therefore, we conclude that two opposite-signed convective columnar vortices have a trend of forming a stable pair and exhibit correlated motions.

\subsection{Collective Motion}

\begin{figure}
  \centerline{\includegraphics[width=1.0\textwidth]{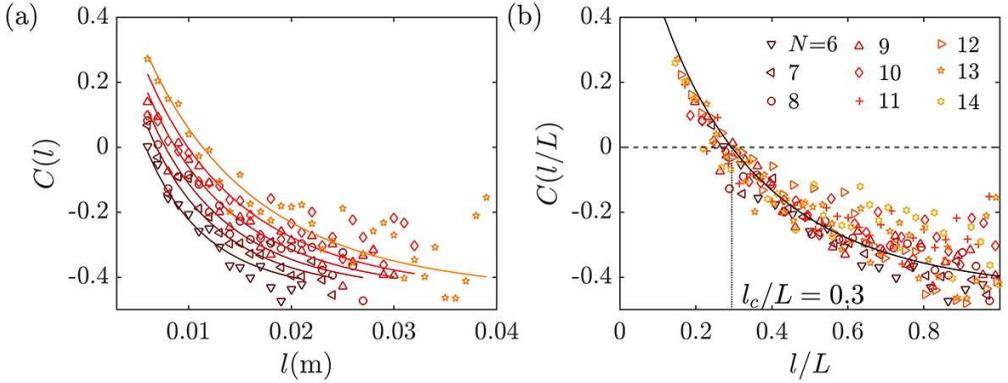}}
  \caption{(a) The spatial correlation function of vortex velocity fluctuation within a cluster $C(l)$ as a function of vortex distance $l$. The solid curves represent the fitted curves $C(l){=}(1.45)e^{{({-}l/H/{c})}^{0.85}}{-}0.45$. The fitted parameter $c$ for various $N$ is ($c{=}0.077$, $N{=}6$), ($c{=}0.091$, $N{=}7$), ($c{=}0.100$, $N{=}8$) ($c{=}0.115$, $N{=}9$), ($c{=}0.131$, $N{=}10$) and ($c{=}0.150$, $N{=}13$). Data of $C(l)$ for larger $N$ appear more scattered about the fitted curve. (b) The correlation function $C(l/L)$ as a function of $l/L$. The vertical dashed line indicates the correlation length $l_{c}{\approx}0.3L$ determined by the zero-crossing position of $C(l/L)$. Data are for $Ra{=}3.0{\times}10^7$ and $Ra/Ra_{c}{=}1.97$. 
%(c) Plot of $l_c/L$ versus $Ra/Ra_c$ for various $Ra$ numbers. The dash line shows the mean value of $l_c/L$ of all data, which is 0.28.
  }
\label{fig:colmo}
\end{figure}

We observe that densely distributed vortices are organized into clusters and move collectively. To further analyze their collective motion, we calculated the spatial correlation function of vortex velocity fluctuation within a cluster \\ $C(l){=}\sum_{ij}[\vec{u}_i'(\vec{r}_i{+}l){\cdot}\vec{u}_j'(\vec{r}_j)\delta(l{-}l_{ij})]/[C_0\sum_{ij}\delta(l{-}l_{ij})]$, where $\vec{u}_i'{=}\vec{u}_i{-}\vec{V}$ is the relative vortex velocity with respect to the mean cluster velocity $\vec{V}{=}\sum_i \vec{u}_{i}/N$, $l_{ij}$ is the distance between the vortex pair $(i,j)$ and $C_0$ is a normalization constant. $\delta(l{-}l_{ij})$ is a Dirac function selecting pairs of vortices separated by distance $l$. Figure 10a shows that $C(l)$ decreases as the distance $l$ increases, with a larger decay length for larger clusters. Figure 10b presents the correlation function $C(l/L)$ as a function of $l$ scaled by the cluster length $L$ for $Ra{=}3.0{\times}10^7$ and $Ra/Ra_{c}{=}1.97$. We see that for clusters with various sizes $N$, $C(l/L)$ collapses approximately onto a single stretched exponential function
\begin{equation}
C(l/L)=1.45{e}^{{(-l/L/0.245)}^{0.85}}-0.45,
\label{eq:cl}
\end{equation}
which crosses zero at the correlation length $l_{\mathrm c} \approx 0.3L$ for all cluster sizes. 

\begin{figure}
	\centerline{\includegraphics[width=0.6\textwidth]{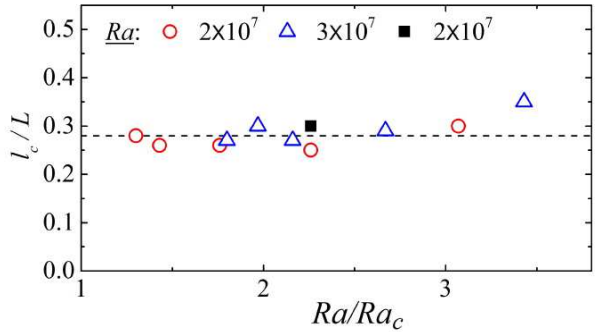}}
	\caption{Plot of $l_c/L$ versus $Ra/Ra_c$ for various $Ra$ numbers. Open circles: experimental data for $Ra{=}2.0{\times}10^{7}$. Open triangles: experimental data for $Ra{=}3.0{\times}10^{7}$. Solid squares: numerical data for $Ra{=}2.0{\times}10^{7}$. The dashed line shows the mean value $\langle{l_c}/L\rangle{=}0.28$ for all data}
	\label{fig:colmo}
\end{figure}

Results of  $l_{c}/L$ are shown as a function of $Ra/Ra_{c}$ in figure 11 for various $Ra$. We see indeed over the parameter range studied $l_{c}/L{=}0.28{\pm}0.03$ remains a constant independent of $Ra$ and $Ra/Ra_{c}$. Since $l_{c}$ represents the characteristic spatial range in which the vortex movements remain correlated, the fact that $l_{c}/L$  is significantly larger than individual vortex size and independent of the cluster size suggest that the correlated motions of the vortices are long-range and scale-free. Scale-free collective behaviors are also observed in many non-equilibrium dynamical systems consisting of densely distributed, interacting entities, e.g. bird flocks, bacteria swarms and active matters \citep{CCGPSSV10,CDBSZ12,WWG14}. For instances, \cite{CCGPSSV10} observed scale-invariant, correlated motions in starling flocks. \cite{CDBSZ12} found that the correlation length of bacterial motion in clusters is approximately $30\%$ of the spatial size of the clusters.
The appearance of scale-free motion of densely populated vortices reported in rapidly rotating convection suggests that it is a general statistical property of collective motion. We remark that the scattering of data points at large distances $(l{\approx}L)$ in figure 10b, owing to insufficient statistics, has negligible influence in the determination of  $l_{\mathrm c}$. 

\subsection{Centrifugal Motion of Clustered Vortices}

\begin{figure}
 \centerline{\includegraphics[width=0.9\textwidth]{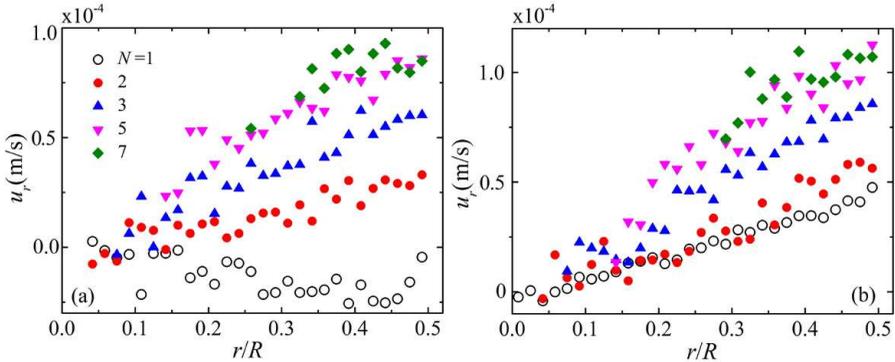}}
  \caption{Ensemble-averaged radial velocity ${\langle u_r\rangle_{\xi}}$ of the vortices as a function of $r/R$ for various cluster sizes $N$. Data are for cyclones (a) and anticyclones (b).  Results for $Ra{=}2.0{\times}10^7$ and $Ra/Ra_{c}{=}2.26$.}
\label{fig:cycur}
\end{figure}

We have shown that it is the interaction between adjacent opposite-sign vortices that organizes individual vortices to form vortex clusters. Within each vortex cluster, the translational motion of the vortices are closely correlated and restricted with each other. In the following we make comparative studies of the radial motions of clustered vortices (inside clusters) and isolated vortices (outside clusters). 

Figure 12 shows the ensemble-averaged radial velocity $\langle u_r\rangle_{\xi}$ for the vortices in clusters with various size $N$. For isolated cyclones that moves individually (i.e., $N{=}1$), $\langle u_r\rangle_{\xi}$ is negative and its magnitude increases approximately linearly with $r$, indicating the normal inward motion of cyclones. However, for clustered cyclones ($N{\ge}2$), $\langle u_r\rangle_{\xi}$ becomes positive which implies anomalous outward motion. We see that the slope of $\langle u_r\rangle_{\xi}(r)$ increases when $N$ increases up to 5. Therefore, clustered cyclones gain larger velocity of inverse centrifugal motion when the cluster size increases. 
The velocity profile of anticyclones exhibits a similar cluster-size dependence as the cyclones, except that for $N{=}1$ $\langle u_r\rangle_{\xi}(r)$ is positive since isolated anticyclones move outwardly.

We determine the slope $\lambda_u$ of the velocity profile, through linear fitting of the data $\langle u_r\rangle_{\xi}$  in the region ($0{\le}r{\le}0.5R$). Results of $\lambda_u$ as a function of cluster size $N$ for $Ra/Ra_c{=}2.67$ are shown in figure 13a. We see clearly that in this inverse-centrifugal regime $\lambda_u$ for cyclones changes sign when $N$ exceeds one. Data for both types of vortices suggest a similar trend for $N{\ge}2$, i. e., $\lambda_u$ increases with $N$ for small cluster sizes and reaches a maximum at $N{=}5$. We also note that for small cluster size $\lambda_u$ is considerably larger for anticyclones than cyclones. 
%The difference of $\lambda_u$ for the two types of vortices becomes minor when $N{\ge}5$. 

When vortices inside a cluster move collectively, they share a similar radial velocity profile  ${\langle u_r\rangle_{\xi}}(r)$. It is thus reasonable to consider all vortices inside the cluster as a single structure that undergoes radial motion driven by the centrifugal force. We propose a Langevin-type equation analogous to (4.3) to describe, to some extent, the radial motion of vortex clusters, whereby the relaxation time $t_{c}{=}M/\eta$ for cluster motion is given by the inertia mass $M$ of all vortices inside the cluster, and ${\zeta}r$ defines the net centrifugal force for all vortices. Since in the inverse-centrifugal regime both the population and vorticity strength of anticyclones overrides that of cyclones, the net centrifugal force is positive and thus the vortex cluster moves outwardly. The solution of the radial velocity profile can be obtained through similar a derivation presented in section 4.2, which has the same form as equation (4.5). For a given initial position of the cluster, one finds ${\langle u_r\rangle_{\xi}}{=}{\lambda_u}r$ with the mobility ${\lambda_u}{=}\sqrt{1/({4t_c}^2){+}\zeta}{-}1/(2t_c)$. 

\begin{figure}
 \centerline{\includegraphics[width=0.9\textwidth]{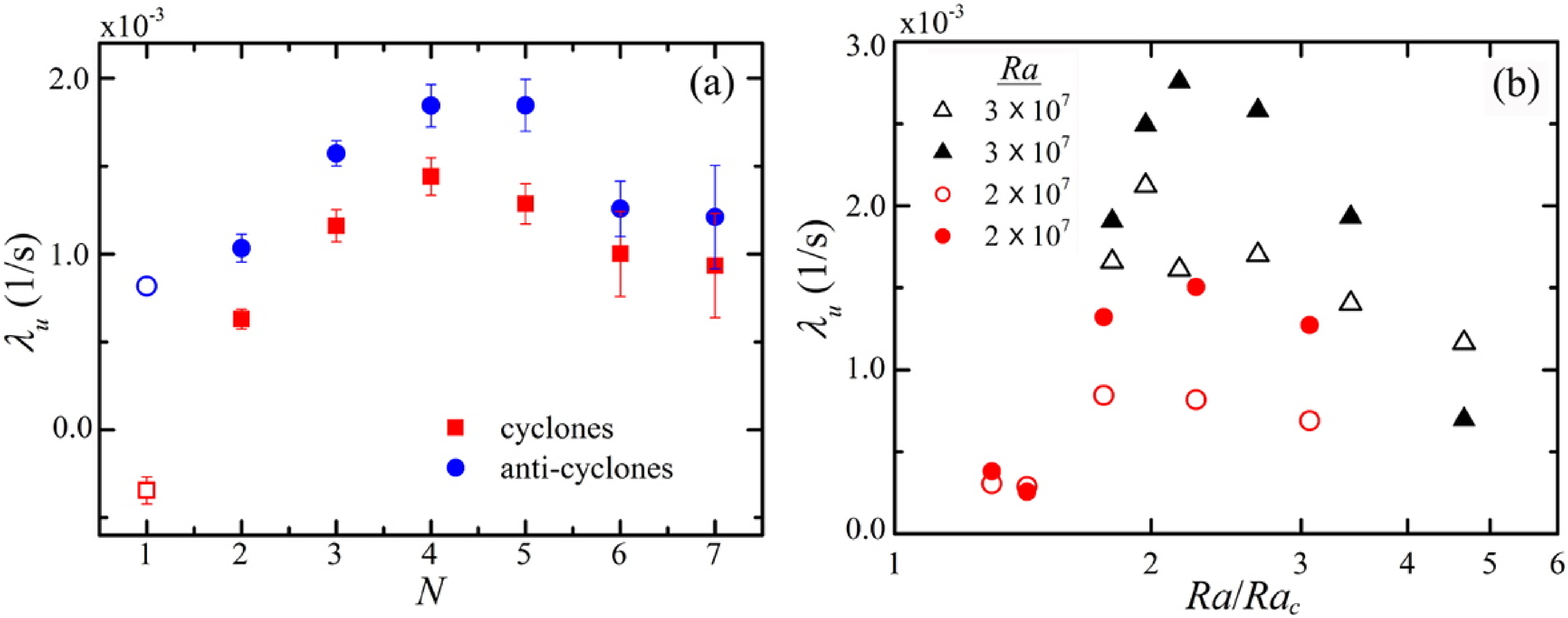}}
  \caption{Mobility $\lambda_u$ of vortex motion determined from the slope of ${\langle u_r\rangle_{\xi}}(r)$. (a) $\lambda_u$ as a function of $N$ for $Ra{=}2.0{\times}10^7$ and $Ra/Ra_{c}{=}2.26$. (b) $\lambda_u$ as a function of $Ra/Ra_{\mathrm c}$ for various $Ra$. Open symbols are data for isolated anticyclones ($N{=}1$). Solid symbols are data for clustered anticyclones ($N{>}1$).}
\label{fig:mobility}
\end{figure}

As shown in figure 13a the mobility of clusters ${\lambda_u}$ is in the order of $10^{{-3}}(1/s)$, and $t_{c}{\approx}10s$ in the inverse-centrifugal regime~\citep[see][]{Paper2}, we find $1/(2t_c){\gg}\lambda_u$, $\zeta{\ll}1/(2t_c)^2$ and thus $\lambda_u{\approx}{\zeta}t_c{=}\Omega^{2}{\langle}{\delta}T{\rangle}t_{c}{=}\Omega^{2}{\langle}{\delta}T{\rangle}M/\eta$. Here ${\langle}{\delta}T{\rangle}$ is the averaged temperature anomaly of the vortex cluster. With increasing cluster size $N$, the inertia mass $M$ of the cluster increases. We suggest that it is the dominating factor for an enlarged ${\lambda_u}$ for clusters of intermediate size ($N{\approx}5$). For very large cluster size ($N{\ge}5$) we infer that viscous damping may become significant to influence cluster motion and thereby $\lambda_u$ decreases.

% Since the temperature anomaly ${\delta}T$ of the convective vortices represents the buoyancy forcing, which is predicted to be proportional to the vorticity magnitude of the vortices \citep{PKVM08, GJWK10}, the mobility of clusters approximates to $\lambda_u{\approx}c^*(z)\Omega^2{\langle}\omega{\rangle}M/\eta$. Here ${\langle}\omega{\rangle}$ is the averaged vorticity over all vortices inside a cluster. $c^{*}(z)$ is a fluid-depth dependent coefficient.**********

Figure 13b presents $\lambda_u$ as a function of $Ra/Ra_{c}$ for isolated ($N{=}1$, open symbols) and clustered ($N{>}1$, solid symbols) anticyclones. Data for two sets of $Ra$ are shown for comparisons. One sees that for a given $Ra$ and $Ra/Ra_{c}$, $\lambda_u$ is greater for clustered anticyclones than for isolated anticyclones. 
For all cases $\lambda_u$ has a maximum at $Ra/Ra_{c}{\approx}2$. We  see that the maximum $\lambda_u$ for clustered anticyclones is nearly twice that of the isolated ones for both $Ra$ numbers. We also see that $\lambda_u$ increases further when a larger $Ra$ is chosen. These results demonstrate that the radial motion of anticyclones are enhanced by the collective motion to a great extent. 

\begin{figure}
	\centerline{\includegraphics[width=0.9\textwidth]{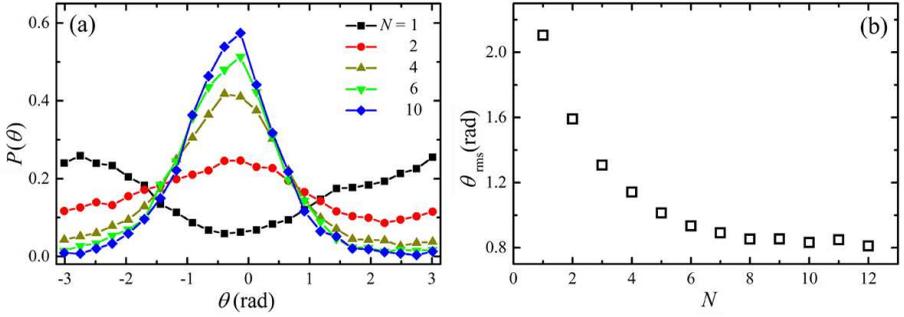}}
	\caption{ (a) Probability density functions of $\theta$ of cyclones in clusters with various size $N$. (b) The standard deviation of $\theta$ as a function of the cluster size $N$. Results for ${Ra}{=}3.0{\times}10^7$, ${Ra}/{Ra_c}{=}1.97$ and ${Fr}{=}0.27$}
	\label{fig:mobility}
\end{figure}

We further investigate the direction of motion $\hat{u}$ of cyclones. Denoting the angle $\theta$ between $\hat{u}$ and the vortex position $\hat{r}$, we show in figure 14a the probability density function $p(\theta)$ for cyclones in clusters with various sizes.  We find that $\theta$ depends strongly on the cluster size $N$. For isolated cyclones ($N{=}1$), the most probable direction of motion is radially inward ($\theta_\mathrm{p}{=}\pi$). However, for clustered cyclones ($N{\ge}2$) we find $\theta_\mathrm{p}{=}0$ as they move outward. For small cluster size $\theta$ is widely distributed and thus $p(\theta)$ exhibits a broad half width. When the cluster size increases we find the distribution of $\theta$ becomes concentrated around $\theta_\mathrm{p}{=}0$ with increasing maximum of $P(\theta)$. Figure 14b shows that the standard deviation of $\theta$ indeed decreases monotonically when $N$ increases. With increasing cluster size $N$ the inertial effect of the vortex cluster becomes significant. Our data reveal that the inverse centrifugal motion of cyclones appears more unidirectional.

\subsection{Froude-number Dependence of the Vortex Mobility}

The mobility $\lambda_{u}$ represents the primary features of radial motion for both isolated and clustered vortices. 
We suggest that the non-monotonic dependence of $\lambda_{u}$ on $Ra/Ra_{c}$ shown in figure 13b is thus owing to the decreasing vorticity magnitude ${\vert}\omega{\vert}$ that competes with the increasing rotation rate $\Omega$. 
Following discussions in sections 4.2 and 6.3, we have $\lambda_u{\approx}{\zeta}t_c{=}\Omega^{2}{\delta}Tm/\eta$. The temperature anomaly ${\delta}T$ of the convective vortices represents the buoyancy forcing, which is predicted to be proportional to the vorticity magnitude of the vortices \citep{PKVM08, GJWK10}. Thus the mobility of clusters can approximate to $\lambda_u{\approx}c^*(z)\Omega^2{\vert}\omega{\vert}m/\eta$, with the coefficient $c^*(z)$ depending on the fluid depth. This approximate relation implies that the mobility of the vortices is determined not only by the rotating rate but also the vorticity magnitude of the vortices.
In figure 15a we show the scaled mobility $\lambda_a/{\vert}\omega{\vert}$ as a function of $Fr$ for both types of vortices. Here $\lambda_a$ is determined by linear fitting of the radial velocity profile ${\langle u_r\rangle_{\xi}}(r)$ shown in figures 12 and 13 for both isolated ($N{=}1$) and clustered vortices ($N{>}1$). 
Since $\lambda_a/{\vert}\omega{\vert}{=}c^{*}(z)\Omega^2m/\eta$, the scaled mobility 
%reveals the relative rate of vorticity transport which 
is expected to increases when the rotation rate increases. Our experimental results in figure 15a show that it is the case as $\lambda_a/{\vert}\omega{\vert}$ for both cyclones and anticyclones grows rapidly with increasing $Fr$ in the full parameter range studied. Meanwhile we see that results for $\lambda_a/{\vert}\omega{\vert}$ are scattered and there exists notable difference between the cyclonic and anticyclonic data for given $Fr$ and $Ra$. 
%$\lambda_a/{\vert}\omega{\vert}(Fr)$ appears to be dependent on $Ra$.

\begin{figure}
	\centerline{\includegraphics[width=1.0\textwidth]{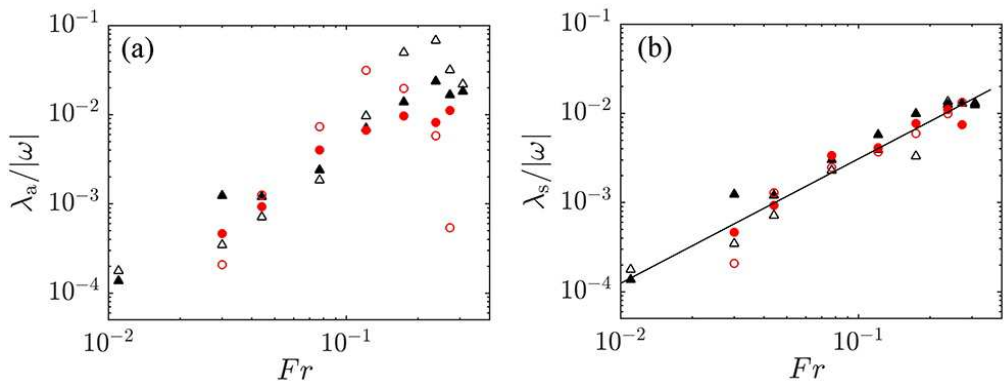}}
	\caption{ (a) The mobility $\lambda_{a}/|\omega|$ of all vortices scaled by the vorticity magnitude as a function of $Fr$. (b) The scaled mobility $\lambda_{s}/|\omega|$ of isolated vortices as a function of $Fr$. Results are for $Ra{=}2.0{\times}10^7$ (red circles) and $Ra{=}3.0{\times}10^7$ (black triangles). Open symbols: cyclones. Solid symbols: anticyclones. The solid line represents a power-function fit with an exponent of $1.40{\pm}0.13$.}
	\label{fig:mobility}
\end{figure}

We suggest that the vortex-interaction of vortices forming clusters may disturb individual vortex motion and modify the vortex mobility. In supporting this argument, we determine the mobility $\lambda_{s}$ for isolated vortices, calculating the slope of the velocity profile ${\langle u_r\rangle_{\xi}}(r)$ (e. g. shown in figure 13) for both cyclones and anticyclones for the case of $N{=}1$. Results of $\lambda_s/{\vert}\omega{\vert}$ are shown as a function of $Fr$ in figure 15b. We see that the discrepancy of $\lambda_s/{\vert}\omega{\vert}$ between isolated cyclones and anticyclones is minor. Moreover, data of $\lambda_s/{\vert}\omega{\vert}$ for various $Ra$ collapse onto a power-low function $\lambda_s/{\vert}\omega{\vert}{\propto}Fr^{1.40{\pm}0.13}$ in the full parameter range of $Fr$ studied, suggesting a general scaling relationship. These results imply that the mobility of isolated vortices scaled by the magnitude of their vorticity can be adequately determined by the Froude number.

% It should be mentioned that for cyclones $\lambda^s=\lambda^*$ since $\lambda^*$ are obtained through fitting into the data of cyclones moving alone in the normal region ($r_c<r<0.9R$). Caption of Fig. 15 

%\input{7-Conclusions}\label{sec:7}

\section{Summary and Discussions}\label{sec:8}
%vortex dynamics.  centrfigual effect -> radial motion -> model,   cluster->collective-> cyclones and anticyclones, formation area->shielding layer. OBTAIN THAT mobility scaling.

A major emphasis of previous studies in rotating convection has been on the flow structures of the columnar vortices \citep{VE02, SJKW06, PKVM08, GJWK10, NRJ14, RKC17, SLDZ20}. As reported briefly in earlier works, these columnar vortices exhibit intriguing translational horizontal motions \citep{Sa97, KA12, NTYM19, Paper1, Paper2}. We present here experimental and numerical studies of the vortex dynamics in rotating RBC in the parameter range of  $2.0{\times}10^7{\le}Ra{\le}2.7{\times}10^8$,  $1.7{\times}10^{-5}{\le}Ek{\le}2.7{\times}10^{-4}$ and $0{\le}Fr{\le}0.31$, which covers the four flow regimes of vortex motion, (I) random vortex diffusion, (II) centrifugal-forced diffusion, (III) inverse-centrifugal motion and (IV) asymptotic centrifugal motion. We found that under slow rotations with $Ra{\gtrsim}10Ra_{c}$, the vortices undergo Brownian-type random motion, i.e., the MSD of the vortices first increases in time as $t^2$ in the ballistic regime, and then linearly in the diffusive regime. Our close inspection of the PDFs of the vortex displacements reveal that, however, the diffusive motion of the vortices is non-Gaussian at small time intervals.
%indicating that the stochastic vortex displacement is correlated to some extent in short time scales.
Furthermore, we show that in this weakly rotating convection state, the vortices are randomly distributed over the horizontal plane, with the PDFs of the Voronoi cell areas of the vortices well described by the standard $\Gamma$ distribution. 

With modest strength of rotations the centrifugal force influences the dynamics of the columnar vortices. We observed that in this flow regime, cyclones move radially towards the rotation axis while anticyclones migrate outwards. Such radial motions of the vortices lead to their super-diffusive behavior, i. e. , the MSD of the vortices increases faster than a linear function of time at large time. Space-time plots of the radial trajectories of the vortices suggest that despite their significant fluctuations, the mean radial displacements of the vortices follow approximately exponential functions of time. Moreover, we note that the ensemble-averaged of the radial velocity $\langle u_r\rangle_\xi$ of cyclones (anticyclones) is negative (positive) and increases (decreases) linearly with respect to $r$. Based on these observations, we proposed an extended Langevin model incorporating the centrifugal force to interpret the vortex motion in the centrifugation-influenced flow regime. The model provides predictions of the first and second moments of the radial displacements of both types of the vortices which coincide with the experimental data, and explains the linear radial dependence of $\langle u_r\rangle_\xi$. Our model thus presents essential interpretations of the super-diffusive behavior of the horizontal vortex motion observed in rotating convection systems \citep{NTYM19, Paper2}.

% In these two regimes, cyclones are dominant vortices since their vorticity magnitude and density are larger than those of anticyclones. 

With increasing rotation speed the horizontal scale of the columnar vortices decreases. It is found that the mean area of the Voronoi cells of the vortices decreases significantly in the flow regime with $1.6Ra_{c}{\le}Ra{\le}4Ra_{c}$, signifying that the vortices are densely distributed. The PDFs of the Voronoi cell area deviate markedly from the standard $\Gamma$ distribution. In this convection state, the hydrodynamic interaction of neighboring vortices becomes prominent to influence the vortex dynamics. We report the existence of a most probable separation between two adjacent counter-rotating vortices that tends to form a vortex pair and move collectively. Such vortex interactions eventually lead to the formation of large-scale vortex clusters. We show that inside the clusters the vortices exhibit correlated motion. The correlation of the translation velocity fluctuations of the vortices is scale invariant, with the correlation length being approximately $30\%$ of the length scale of the clusters. In this centrifugation-dominated flow regime, the cold anticyclones override the warm cyclones both in vorticity strength and population, a flow-field asymmetry brought about by the centrifugal effect. Within vortex clusters the motion of the weak cyclones thus submit to that of strong anticyclones and move outwardly in a collective manner. We suggest that this is the underlying mechanism for the counterintuitive inverse-centrifugal motion of the cyclones. 

Within each vortex clusters, the translational motion of the vortices are long-range correlated. We show that such correlated motion of the clustered vortices exerts essential influences on their dynamics. With increasing cluster size $N$, the radial velocity $\langle u_r\rangle_{\xi} (r)$ for clustered cyclones and anticyclones increases faster with $r$, thus both types of vortices gain a larger translation velocity moving outwardly. We also note that the translational motion of cyclones in larger clusters is more concentrated in the outward direction, indicating that their inverse centrifugal motion becomes more unidirectional. 
Finally, despite the complicated $Fr$-dependence of the scale vortex mobility $\lambda_a/{\vert}\omega{\vert}$ for all vortices, we discover in the full parameter range of $Fr$ a simple power-law scaling $\lambda_s/{\vert}\omega{\vert}{\propto}Fr^{1.4}$ for isolated vortices for various $Ra$ and vortex types.  

We have shown in this work the rich and intriguing vortex dynamics in rotating RBC. There remain numerous issues requiring further investigations. Prominent among those include the origin of the Brownian but non-Gaussian diffusion of the vortex displacements. Such non-Gaussian diffusive motion becomes apparent at short time scales. Another crucial issue awaiting theoretical descriptions is the hydrodynamic interactions among closely located convective vortices that give rise to the long-range collective motion of the vortices. Equally important would be explanations of the scale-free correlation of the vortex velocity fluctuations that exist in large-scale vortex clusters. What are the implications of the vortex dynamics presented in this study for the correlated vortex motion observed in giant gaseous planets and other large-scale geophysical and astrophysical flows \citep{LIKB20,MABMGP21,GK21,LIKB22}? We expect future progress in answering these questions.

\backsection[Acknowledgements]{This work is supported by the National Science Foundation of China under Grant No. 92152105, 12232010 and 12072144, a NSFC/RGC Joint Research Grant No. 11561161004 (JQZ) and N\_CUHK437$/$15 (KQX) and by the Hong Kong Research Grants Council under Grant No. 14301115 and 14302317. JQZ acknowledges the support from the Research Program of Science and Technology Commission of Shanghai Municipality.}
%{This work is supported by the Fundamental Research Funds for the Central Universities, and the National Science Foundation of China under Grant Nos. 92152105 and 11772235.}
\backsection[Author contributions]{J.-Q.Z. and K.-Q.X. conceived and designed research. S-S.D. and W.-T.W. conducted the experiments. G.-Y.D. and K.L.C conducted the numerical simulations. S-S.D., J.-Q.Z. and K.-Q.X. wrote the manuscript.}
\backsection[Declaration of interests]{The authors report no conflict of interest.}

\bibliographystyle{jfm}
%\bibliography{jfm2esam}

\bibliography{refs_long}

\end{document}